\documentclass[5p,10pt]{elsarticle}

\usepackage{mathtools}
\usepackage{lineno}
\modulolinenumbers[5]
\usepackage{booktabs}
\usepackage{multirow}
\usepackage[table, dvipsnames]{xcolor}
\usepackage{graphicx}
\usepackage{graphics}
\usepackage{boldline} 
\usepackage{array}
\usepackage{caption}
\usepackage[export]{adjustbox}
\usepackage[colorlinks]{hyperref}
\usepackage{longtable}
\usepackage{tabularray}
\usepackage{makecell}
\usepackage{xcolor}
\usepackage{float}
\usepackage[export]{adjustbox}
\usepackage{geometry}
\usepackage{gensymb}
\usepackage{eurosym}
\usepackage{xurl}
\usepackage[resetlabels,labeled]{multibib}
\newcites{S}{Supplementary References}
\biboptions{numbers,sort&compress,super}

\journal{Springer}

\newcommand{\beginsupplement}{%
	\setcounter{table}{0}
	\renewcommand{\thetable}{S\arabic{table}}%
	\setcounter{figure}{0}
	\renewcommand{\thefigure}{S\arabic{figure}}%
	\setcounter{section}{0}
	\renewcommand{\thesection}{Supplemental \arabic{section}}%
	\setcounter{equation}{0}
	\renewcommand{\theequation}{S\arabic{equation}}
	\setcounter{page}{1}
}

\begin{document}

\begin{frontmatter}

\title{Strategic deployment of solar photovoltaics for achieving self-sufficiency in Europe throughout the energy transition
}

\author[mpe]{Parisa Rahdan\corref{cor1}  }
\cortext[cor1]{Lead contact and corresponding author, Email: parisr@mpe.au.dk}
\author[TUB]{Elisabeth Zeyen}   
\author[mpe,novo]{Marta Victoria}

\affiliation[mpe]{organization={Department of Mechanical and Production Engineering and iCLIMATE Interdisciplinary Centre for Climate Change, Aarhus University},
            addressline={}, 
            postcode={8000}, 
            city={Aarhus},
            country={Denmark}}

\affiliation[TUB]{organization={Department of Digital Transformation in Energy Systems, Technische Universität Berlin Einsteinufer 25 (TA 8)},
            postcode={10587}, 
            city={Berlin},
            country={Germany}}     
        
\affiliation[novo]{organization={Novo Nordisk Foundation $CO_2$ Research Center},
            addressline={Gustav Wieds Vej 10}, 
            postcode={8000}, 
            city={Aarhus},
            country={Denmark}}

\begin{abstract}
 
Transition pathways for Europe to achieve carbon neutrality emphasize the need for a massive deployment of solar and wind energy. Global cost optimization would lead to installing most of the renewable capacity in a few resource-rich countries, but policy decisions could prioritize other factors. We investigate the effect of energy independence on Europe’s energy system design. We show that self-sufficiency constraints lead to a more equitable distribution of costs and installed capacities across Europe. However, countries that typically depend on energy imports face cost increases of up to 150\% to achieve complete self-sufficiency. Self-sufficiency particularly favours solar photovoltaic (PV) energy, and with declining PV module prices, alternative configurations like inverter dimensioning and horizontal tracking are beneficial enough to be part of the optimal solution for many countries. Moreover, we found that very large solar and wind annual installation rates are required, but they seem feasible considering recent historical trends.

\end{abstract}


\end{frontmatter}

\section{Introduction}

As Europe advances in its green transition, it is crucial to harmonize national policies to achieve a net-zero emissions system by 2050. This study addresses three major research gaps in macro-energy system planning: the impact of countries’ energy self-sufficiency on transition pathways, the role of new solar photovoltaic (PV) configurations, and the feasibility of high growth rates for wind and solar PV based on historical trends. 

Energy self-sufficiency is the capability to satisfy energy needs without depending on others. Although collaboration will be necessary to achieve a renewable European energy system with minimal costs \cite{brown2018synergies}, most European countries would like to attain a certain degree of self-sufficiency for energy security. This inclination has been intensified by the gas crisis triggered by Russia's invasion of Ukraine, re-opening the discussion in several countries on the convenience of increasing the use of nuclear or even fossil fuels \cite{Ansa_2023,AP_2023,Euractive_2022, bohdanowicz2023support, carfora2022energy, rokicki2023impact} to lessen their reliance on energy imports \cite{UKenergy_2023}. Previous works \cite{neumann2021costs,schwenk2021co2, sasse2023cost, trondle2019home,vangreevenbroek2023enabling} have investigated the role of self-sufficient generation for different European countries, but these analyses were either limited to the power sector, one region, or just one specific point in time, missing the potential path-dependency observed when modeling transition paths. To address these gaps, in this work we examine the effect of self-sufficiency on the transition to a climate-neutral sector-coupled European energy system.

Solar PV electricity is highlighted as the most cost-effective mitigation investment globally \cite{RN15}, and its deployment increases when pursuing regional equity or reducing gas imports for Europe \cite{neumann2021costs,sasse2019distributional,sasse2020regional,pedersen2022long, pedersen2021modeling}. Distributed PV, installed on rooftops or parking lots, can also increase self-sufficiency in highly-populated regions \cite{rahdan2024distributed}. Considering the importance of solar PV both for the transition and achieving self-sufficiency, we aim to improve the representation of solar PV in macro energy systems models. To do this, we model several emerging PV configurations that are often overlooked in studies but could be cost-efficient due to dramatic cost reduction experienced by PV modules, driven by rapid learning curves \cite{victoria2021solar}. First is inverter dimensioning, meaning the inverter converting PV power from DC to AC is undersized on the AC side since PV modules rarely generate power at full capacity (with standard high solar irradiation) \cite{sma_oversizing_2024,fluence_optimize_2024}. This practice lowers costs, even if it leads to some power curtailment during very sunny hours, and is becoming more advantageous as PV module costs decrease faster than AC components like inverter and grid connection. Second, horizontal single-axis tracking (HSAT), where PV modules rotate from east in the morning to west in the evening, extends solar generation hours and is already cost-efficient, holding a 60\% market share in new utility PV installations in 2023-2024 \cite{ITRPV_2022}. Except for the works of Breyer and co-workers \cite{breyer2017role, afanasyeva2018relevance}, HSAT is often excluded from macro-energy system models. Third, inexpensive PV modules enable novel system designs like the delta configuration, which comprises triangular rows of modules facing east and west. This non-optimal orientation results in lower annual energy generation per DC capacity, but matches the daily profile of HSAT without moving parts, increasing self-consumption and attaining higher energy yield per area \cite{chattopadhyay2017impact, zappa2018analysing, sanchez2021exploring, LAHNAOUI20174312}. Other advantages of delta configuration are reduced wind load and structural weight for rooftop systems \cite{mubarak2019pv}, easy installation for the many homes with east or west-facing rooftops \cite{jurasz2020can}, and possible integration with agriculture as vertical east-west bifacial modules  \cite{Agrivoltaics_fraun_2020, ali2023comparative}. In addition to the question of cost-effectiveness, we explore whether these configurations support self-sufficiency for a carbon-neutral Europe.

Installing large-scale capacities of wind and solar PV has been shown to be a cost-effective strategy to achieve a carbon-neutral Europe \cite{victoria_2022_S, victoria2020early,pickering2022diversity,plessmann2017meet,bogdanov2019radical} and increase energy security \cite{pedersen2022long}. However, the large required installation rates for wind and solar have been questioned to be feasible \cite{cherp2021national, hansen2017limits} or possible by social acceptance issues \cite{flachsbarth2021addressing, segreto2020trends, pasqualetti2011social}. We address here the question of whether ensuring a certain degree of self-sufficiency while decarbonizing the different European countries modifies the wind and solar installation rates, and whether these rates can be considered achievable when looking at historical rates. 

This work introduces three main novelties: (i) modeling a Paris Agreement-compatible transition for self-sufficient interconnected European countries, (ii) examining emerging PV configurations, and (iii) assessing required installation rates for wind and solar, contextualized with historical data. We model the European sector-coupled networked energy system from 2025 to 2050 with 5-year steps, under a carbon budget corresponding to 1.7\degree C temperature increase and imposing carbon neutrality by 2050, with a network comprising 37 nodes, using 370 regions to represent wind and solar resources, and a time-resolution of 2-hour for a full year. Results indicate that total system costs change minimally under self-sufficiency, but lead to a fairer distribution of installed capacities. This could increase costs up to 150\% for previously net-importing countries by 2050.  For a 100\% self-sufficiency constraint, high-value synthetic fuels such as methanol and oil are produced in countries with good renewable resources, and hydrogen is traded extensively between countries. Alternative PV configurations, especially reducing inverter capacity and using horizontal single-axis  tracking, are deemed cost-efficient and installed in large capacities, and can help countries reach self-sufficiency by reducing costs and extending solar generation hours. The growth rates for both wind and solar require higher ambitions in many European countries, but the trends of recent years show that they are achievable.

\section{Results and discussion}

\subsection{Self-sufficient European countries can achieve net-zero emissions and energy independence, with only a slight cost increase}

\begin{figure*}[ht!]
	\renewcommand{\figurename}{Fig.}    \includegraphics[width=1\textwidth,]{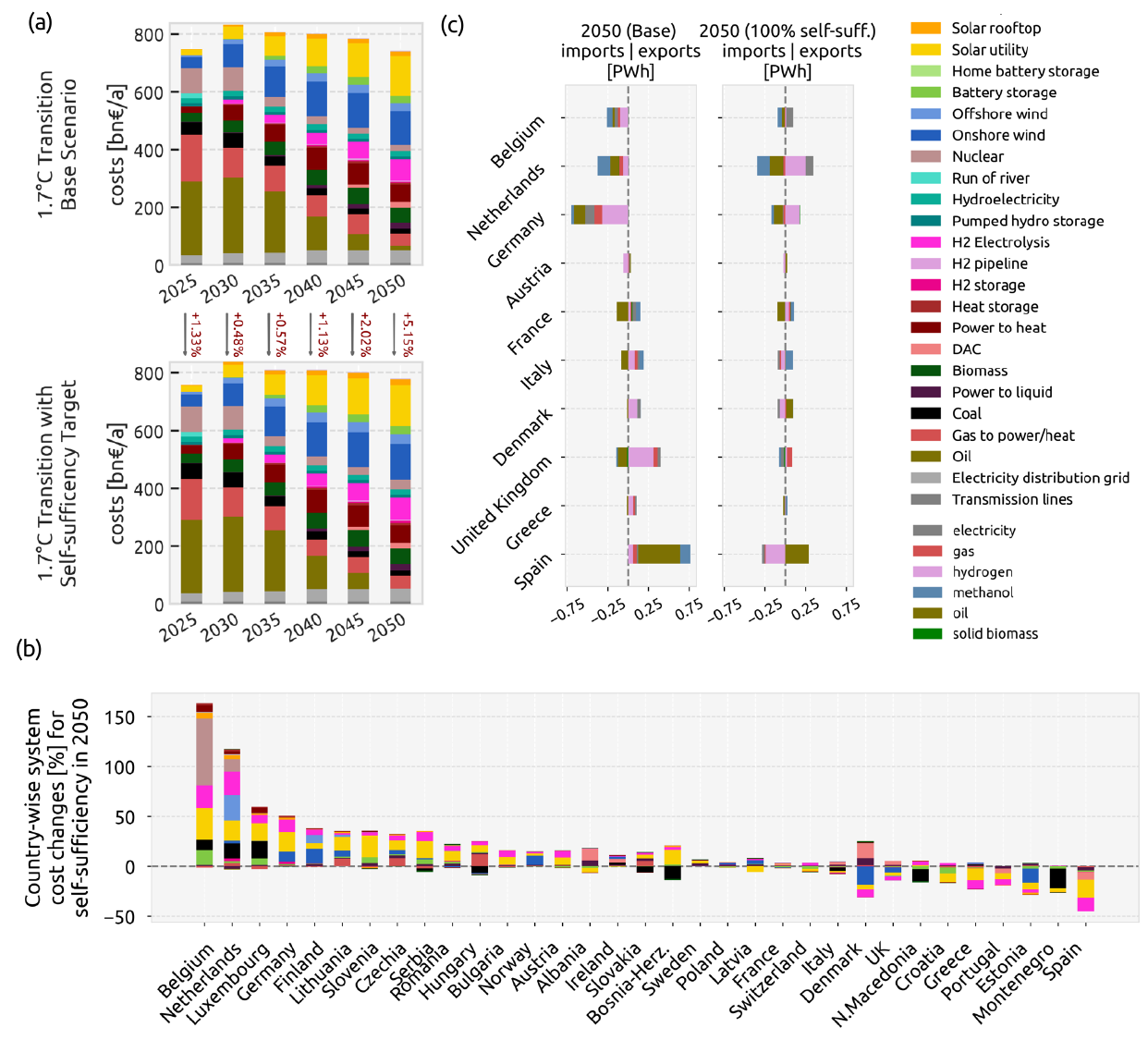}
	\caption{\textbf{} Changes in a) Total system costs during the transition under a 1.7\degree C temperature increase target with and without self-sufficiency, b) country-specific system costs (\%) for year 2050 for the 100\% self-sufficient scenario compared to base scenario, showing what components are responsible for the cost increase (\%) in each country, and c) Energy imports (negative) and energy exports (positive) of each country for year 2050 for the 100\% self-sufficient scenario and base scenario. }
	\label{fig:1}
\end{figure*}

We model the energy transition in Europe from 2025 to 2050 in 5 years time steps, using a myopic approach and a carbon budget corresponding to 1.7\degree C temperature increase (see Supplementary Fig. \ref{fig:S3}). First, we explore the transition path without any self-sufficiency requirements and compare it with the results when adding a self-sufficiency target, which requires that the self-sufficiency coefficient for every country (Eq. \ref{eq:2}) reaches 60\%, 80\%, and 100\% in 2030, 2040, and 2050, respectively. Total system cost composition for both transition paths are shown in Fig. \ref{fig:1}a. Implementing the self-sufficiency constraint increases total system cost by 5.1\% for the last investment period, and shows an average increase of 2.1\% during the whole transition. The small total system cost increase in the case of self-sufficiency indicates that the cost-optimal solution space is very flat (as explored in several studies \cite{sasse2020regional,pedersen2021modeling,grochowicz2023intersecting,pickering2022diversity}), and self-sufficiency could be achieved with small additional expenses.

However, on a national level, the system costs can increase strongly, with some countries experiencing up to a 150\% rise (Fig. \ref{fig:1}b). The increase in total system cost is low because under the self-sufficiency constraint, renewable generation and electrolysis capacities are moved from countries that previously were net-exporters, such as Spain or Denmark, to previous net-importing countries, such as Belgium, Netherlands, or Germany. These countries tend to be net importers due to their high industry demand and population density. For Belgium, which experiences the highest cost increase, the surge is largely driven by the need for additional nuclear capacity when the assumed potential for wind and solar are maxed out, which despite being relatively small, incurs substantial costs due to the high expense of the technology (see Supplementary Fig. \ref{fig:S17}a).

Implementing self-sufficiency results in a more balanced exchange of energy carriers between countries (Fig. \ref{fig:1}c). In the model, energy can be exchanged among countries in the form of electricity by AC and DC transmission lines, biomass by trucks, hydrogen and gas by pipelines, and oil and methanol assuming negligible transport costs and no transmission bottlenecks. Electricity network expansion is limited to 10\% of today's capacity, and only 13\% of the total energy transported between countries is through AC and DC lines. Gas exchange between countries reduces after 2025 (see Supplementary Figs. S14-S17), since gas demand is reduced by electrified heating due to the carbon budget constraint. Consequently, hydrogen, synthetic oil, and synthetic methanol are the most relevant energy imports and exports for the 2050 net-zero emissions system, as they are used in the transport and industry sectors. Hydrogen is mostly produced from electrolysis, and synthetic fuels are produced by combining electrolytic hydrogen and captured CO\textsubscript{2} using Fischer-Tropsch reaction and methanolisation units. 

In the base transition scenario, oil and methanol are either produced in countries with abundant renewable resource, such as Spain, or produced in countries with high demand, such as Germany, by importing the required hydrogen and electricity from neighboring countries with good renewable resource, e.g. Denmark and the UK. However, the self-sufficiency constraint results in spreading the production of synthetic oil, and methanol to a lesser degree, to all countries (Fig. \ref{fig:1}c and Supplementary Fig. \ref{fig:S17}). The share of oil and methanol from total exports is still high in countries such as Spain, the UK, and France. Germany and several other countries import their entire methanol demand, and Netherlands and Portugal rely on imports for over 90\% of their oil demand (see Supplementary Figs. S14-S17). This is because the self-sufficiency constraint is imposed on the sum of all energy carriers (Eq. \ref{eq:3}), which makes the system benefit from producing high-value energy carriers in countries with good renewable resources. Therefore, European cooperation is still necessary for achieving self-sufficiency with minimal costs.

Solar PV and wind energy become the cornerstone of the transformed energy system, with solar PV being crucial for achieving self-sufficiency. By 2050, 5.1 TW of solar and 1.3 TW of onshore and offshore wind capacity are installed across Europe (see Supplementary Fig. \ref{fig:S9}), taking up 57\% and 36\% of the electricity generation, respectively. Solar PV plays a more prominent role in our scenarios than previous similar studies \cite{zeyen2023endogenous,victoria_2022_S}, because we have included inverter dimensioning in the model. Notably, solar PV is the only technology whose capacity consistently increases across all countries that need to boost local generation to achieve self-sufficiency (Fig. \ref{fig:1}b).

\subsection{Solar PV tracking and inverter dimensioning can reduce costs for the green transition}

\begin{figure*}
\renewcommand{\figurename}{Fig.} 
    \includegraphics[width=1\textwidth,]{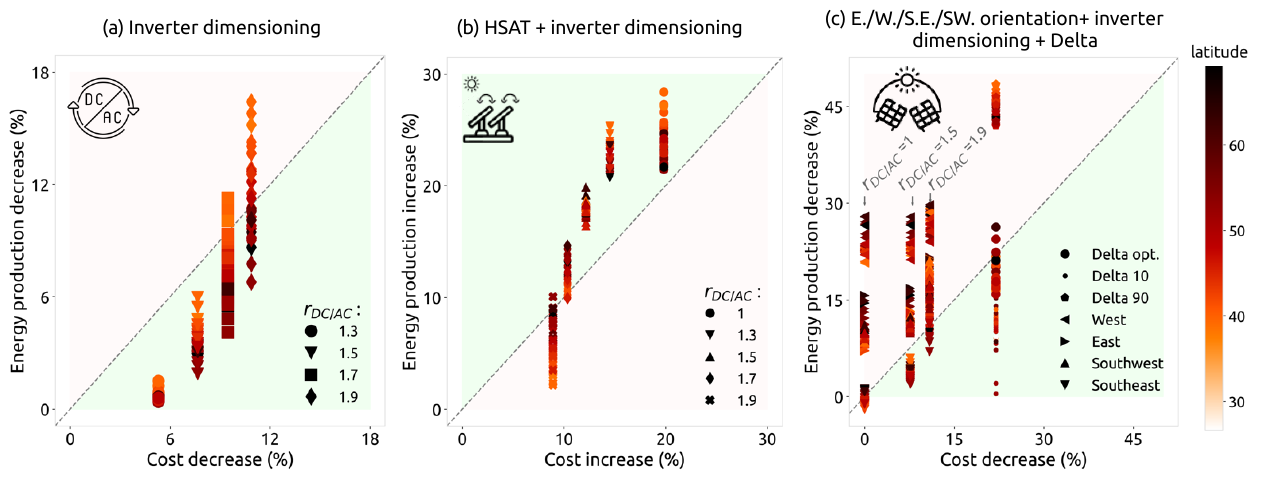}
   \caption{\textbf{} Changes in annual energy generation vs. cost for the different PV configurations compared to a South-facing module with inverter ratio of 1 for: a) higher inverter ratios (\({r}_{DC/AC}\)), b) HSAT with inverter dimensioning, and c) delta configuration, with 10\degree/optimal/90\degree\ inclination, plus east/west/southeast/southwest facing modules with optimal inclination and inverter dimensioning (The optimal inclination is the best tilt angle for south-facing modules in each country). The green areas show configurations that are considered cost-efficient, and the further a point is from the diagonal line, the higher its cost benefit or detriment.}
   \label{fig:2}
\end{figure*}

Given the significant deployment of both utility-scale and distributed solar capacities, we explore the potential of alternative solar technologies in enhancing energy independence. Based on an initial cost-benefit analysis (Fig. \ref{fig:2}), 19 alternative solar PV configurations are selected for evaluation. We then examine two overnight near-zero scenarios, both featuring the 19 configurations, but only one imposing a 100\% self-sufficiency target. For utility-scale PV generation connected to the high-voltage grid, we evaluate three configurations: (i) south-facing and (ii) horizontal single-axis tracking (HSAT), both with different inverter ratios, and (iii) delta configuration with 10\degree\ inclination and 1.5 inverter ratio. For rooftop PV systems connected to the low-voltage grid, we consider the same configurations as utility except HSAT, and instead add southeast and southwest-facing modules with an inverter ratio of 1.9.

Different inverter ratios and HSAT are selected in both overnight scenarios (see Supplementary Fig. \ref{fig:S10}). Delta configuration is only selected for the scenario with 100\% self-sufficiency target. Southwest and southeast are not selected in any case. The reason for a configuration to be selected is mainly based on the energy generation gains vs. the increase in costs (or vice-versa) in every country (Fig. \ref{fig:2}). First, for all countries in Europe, the cost decrease of selecting a DC/AC ratio of 1.3 or 1.5 is higher than the reduction in annual energy generation (green area of the plot in Fig. \ref{fig:2}a). This confirms that solar energy generation at nominal DC capacity is infrequent enough to justify oversizing the PV modules or DC capacity of the plant. For example, a DC/AC ratio of 1.3 results in annual energy loss below 3\% for every country, but higher ratios like 1.7 and 1.9 are not beneficial for Southern Europe due to higher losses. However, cost vs. energy is not the single deciding factor for selection of a configuration, as demand at peak solar production hours, generation from other renewable sources, and the available transmission or storage capacity will also impact the selection.

Second, Fig. \ref{fig:2}b shows all DC/AC ratios except 1.9 are cost-efficient to use with HSAT for all countries in Europe. Interestingly, Southern countries gain the highest increase in energy production for a low DC/AC ratio, but ratios higher than 1.5 are more appropriate for Northern European countries since the number of hours at high DC generation with HSAT is still low (see Supplementary Fig. \ref{fig:S4}). 

Third, east, west, southeast, and southwest-facing modules are generally not cost-efficient, or much less cost-efficient than other configurations. Delta configuration reduces the annual generation (relative to the installed DC capacity, as discussed in Supplementary section S3), but also the cost. Delta configuration with 10\degree\ inclination angle could be beneficial for any European country (Fig. \ref{fig:2}c), but optimal tilt (same as south-facing) is only attractive for higher latitudes where the Sun is low during most of the year.

Despite the results from the simplified evaluation in Fig. \ref{fig:2}c, when alternative orientation angles or delta configurations are evaluated in the overnight cost optimization, they are not selected (see Supplementary Fig. \ref{fig:S9}). This is because HSAT is more cost-efficient than these configurations for increasing PV generation in the early morning or late afternoon. These configurations can however still be useful for single consumers who would benefit from them depending on their demand and the electricity prices in the region.  

Overall, the results here indicate that consideration of inverter dimensioning when modeling solar PV is essential, as for both south-facing fixed modules and HSAT, almost the entire capacity is paired with an inverter ratio higher than 1.

\subsection{Alternative solar PV configurations have benefits for self-sufficiency}

We model again the transition paths including the alternative solar configuration that were found cost-effective in the previous overnight optimization exercise. These include south-oriented configuration with inverter ratios of 1.5 and 1.7 (both for utility-scale and distributed systems), HSAT with inverter ratios of 1.3 and 1.5, and delta configuration with 10\degree\ inclination and inverter ratio of 1.5. 

\begin{figure}[t]
\renewcommand{\figurename}{Fig.} 
    \includegraphics[width=\columnwidth,center]{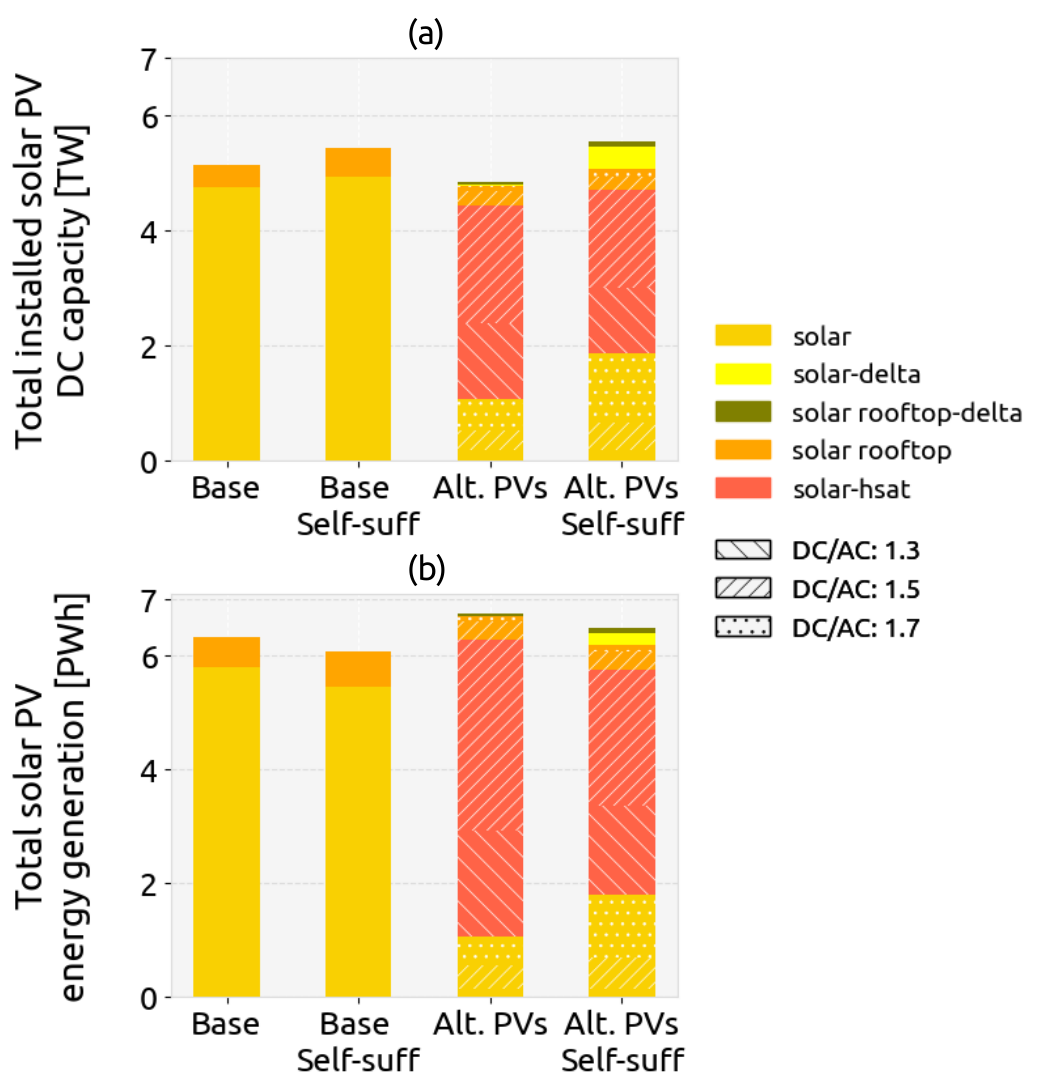}
   \caption{\textbf{} (a) Cumulative installed DC capacity and (b) energy generation from different solar configurations for all transition scenarios in 2050. }
   \label{fig:3}
\end{figure}

\begin{figure*}[!ht]
    \includegraphics[width=1\textwidth,center]{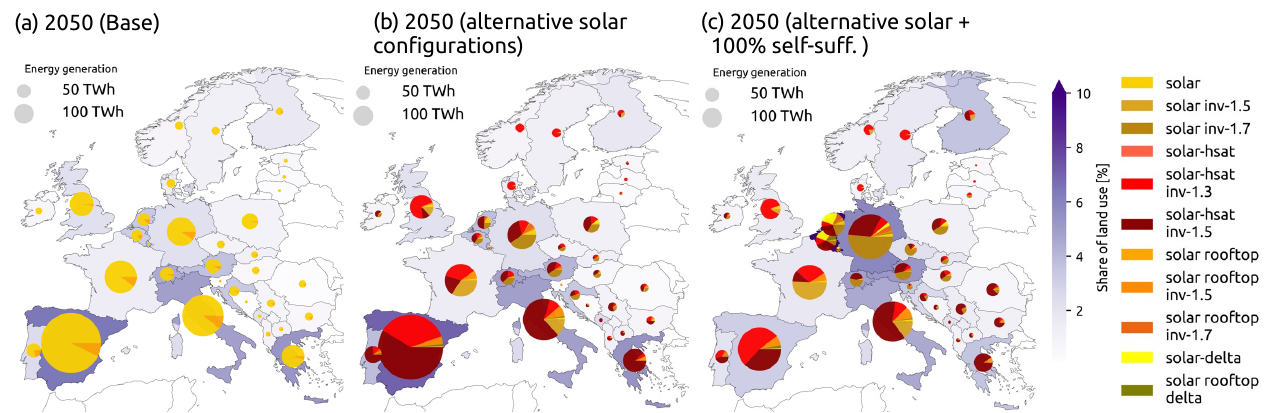}
   \caption{\textbf{} Solar generation map by configuration and regional land-use (\% of available land) of solar PV technologies for year 2050 for a) Base transition b) Transition with selected alternative solar configurations, and c) Transition with selected alternative solar configurations under a 100\% self-sufficiency target.}
   \label{fig:4}
\end{figure*}

Including the self-sufficiency target leads to a lower Europe-average capacity factor of solar PV for both the base scenario and the ``Alternative PV" scenario, since solar PV is installed in less optimal locations (Fig. \ref{fig:3} and Fig. \ref{fig:4}). Therefore, despite higher installed capacity, total solar generation under the self-sufficiency target is decreased (Fig. \ref{fig:3}) and compensated by higher wind generation in the system (see Supplementary Fig. \ref{fig:S9}). Addition of alternative PV configuration results in the installation of large shares of HSAT in many countries. For the ``Alternative PV" scenario, the higher capacity factors attained by HSAT help achieve higher energy generation from solar PV compared to the base scenario despite a lower DC capacity being installed in the system. Under the self-sufficiency constraint, a higher capacity of solar PV is installed in the system when alternative configurations are available. Despite this higher capacity, the total system cost for the transition with self-sufficiency target is lowered by an average of 1.4\% when alternative solar configurations are added (see Supplementary Fig. \ref{fig:S8}). 

HSAT is notably installed in the most southern and most northern countries, with average higher DC/AC ratios in southern European countries (Fig. \ref{fig:4}). These countries benefit from the longer solar production hours that HSAT provides, but do not need the extra generation during peak generation hours around noon (see Supplementary Fig. \ref{fig:S11}), hence the prevalence of 1.5 DC/AC ratio even though 1.3 is more cost-efficient for most countries (Fig. \ref{fig:2}). Adding a self-sufficiency target to the transition has two noticeable impacts when it comes to solar PV installations. First, delta configuration represents a large share of solar generation in Belgium and Netherlands, and a small share in Poland and Germany. Second, there is a shift from HSAT with DC/AC ratio of 1.5 to 1.3 for countries like Spain, France, and Switzerland while Germany and Belgium experience this shift in reverse, installing more HSAT with DC/AC ratio of 1.5. This means that the system still does not require extra generation at noon, and is shifting the capacities previously installed in Spain to other countries. 

\subsection{Large installation rates are within reach}

\begin{figure*}[!ht]
\renewcommand{\figurename}{Fig.} 
    \includegraphics[width=0.95\textwidth,center]{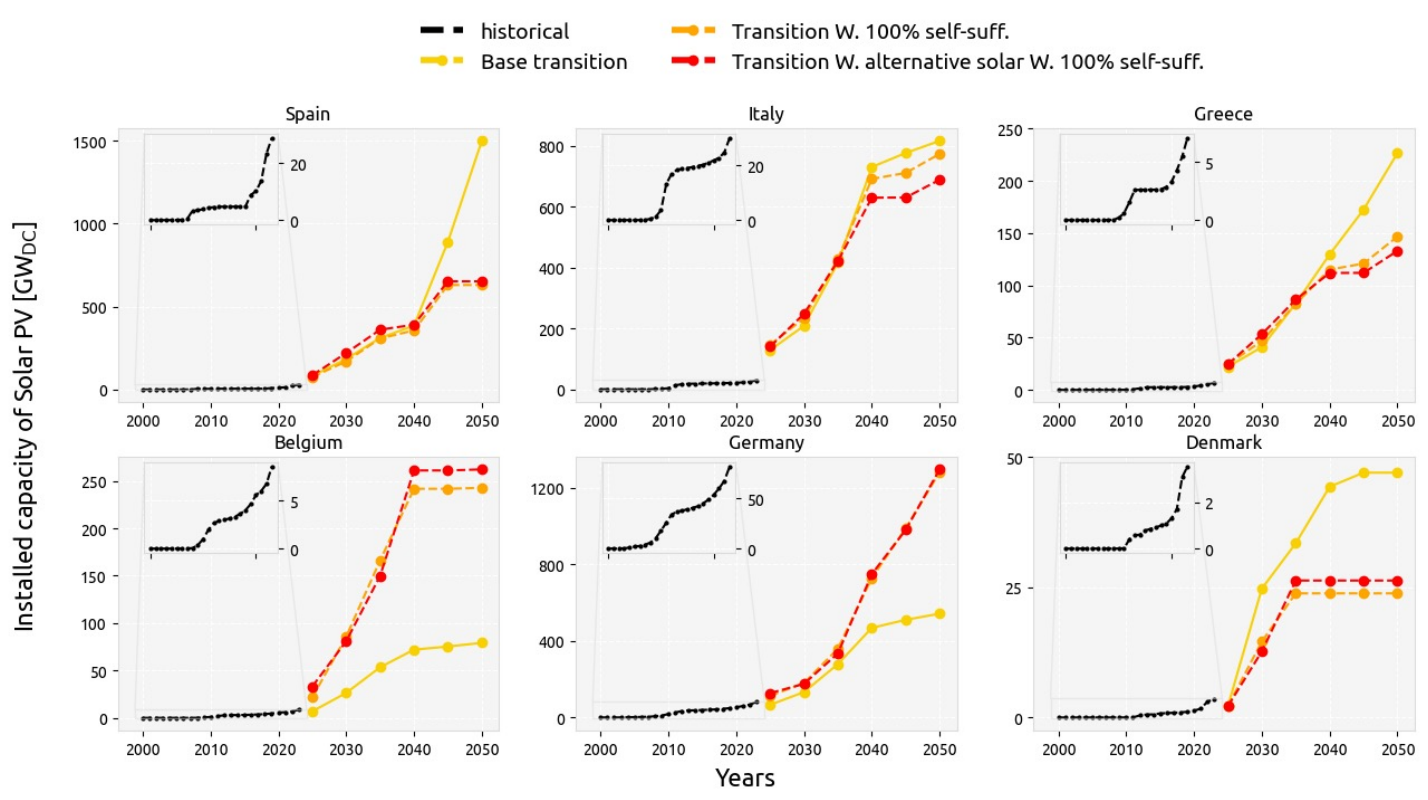}
   \caption{\textbf{} Historical (data from IRENA\cite{IRENA_2022}) and modeled cumulative installed capacity of solar PV for selected countries under base transition, transition with a 100\% self-sufficiency target, and transition with selected alternative solar configurations under a 100\% self-sufficiency target. The insert in each figure shows a zoom on the historical years. Refer to Supplementary Figs. S22 and S23 for other countries and the required installation rates.}
   \label{fig:5}
\end{figure*}

\begin{figure}[H]
\renewcommand*{\thefigure}{S\arabic{figure}} \renewcommand{\figurename}{Fig.} 
\includegraphics[width=1\textwidth, center]{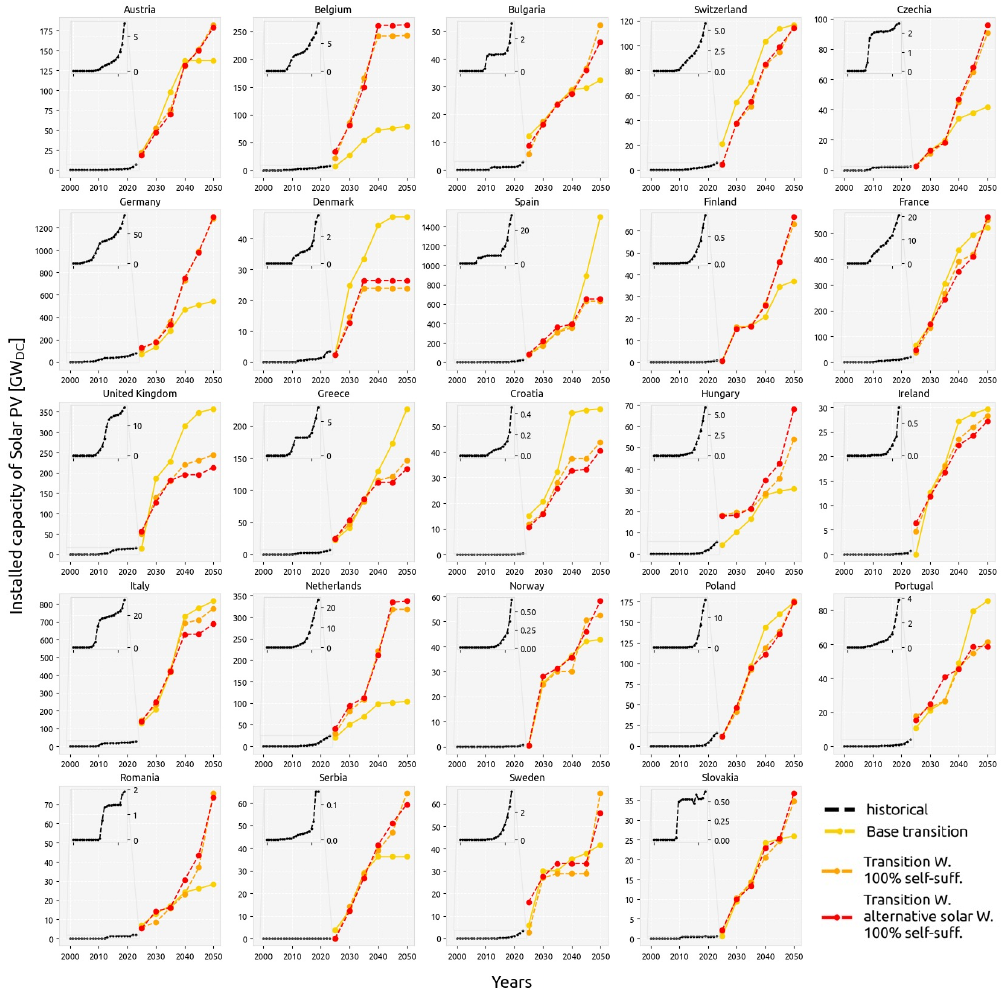}
\caption{Historical (data from IRENA \protect\citeS{IRENA_2022_S}) and future cumulative installed capacity of solar PV for different European countries modeled under base transition, transition with a 100\% self-sufficiency target, and transition with selected alternative solar configurations under a 100\% self-sufficiency target.}
\label{fig:S22}
\end{figure}

Now we turn our attention to how the transition scenarios play out for each country. We compare the historical capacities of solar PV and wind (both onshore and offshore) using data from IRENA\cite{IRENA_2022}, with what is needed from now till 2050. The transition paths of most countries are similar for all scenarios, but the self-sufficiency requirement has a noticeable impact on certain countries (Fig. \ref{fig:5} and Supplementary Fig. \ref{fig:S22}). Countries that were previously net importers, such as Germany and Netherlands, increase their cumulative solar capacity, and the opposite happens in previous net exporters. Including alternative solar configurations could result in both an increase or a decrease in PV capacity, but the change is usually much smaller compared to the impact of the self-sufficiency requirement. One group of countries such as Italy, Portugal, Sweden, and Austria, replace their static PV capacity with HSAT to increase solar generation. The other group, including Spain, Denmark, Belgium, and Netherlands, take advantage of the more cost-efficient configurations and install more solar PV to reach self-sufficiency with lower costs.  

\begin{figure*}[!ht]
\renewcommand{\figurename}{Fig.} 
    \includegraphics[width=0.95\textwidth,center]{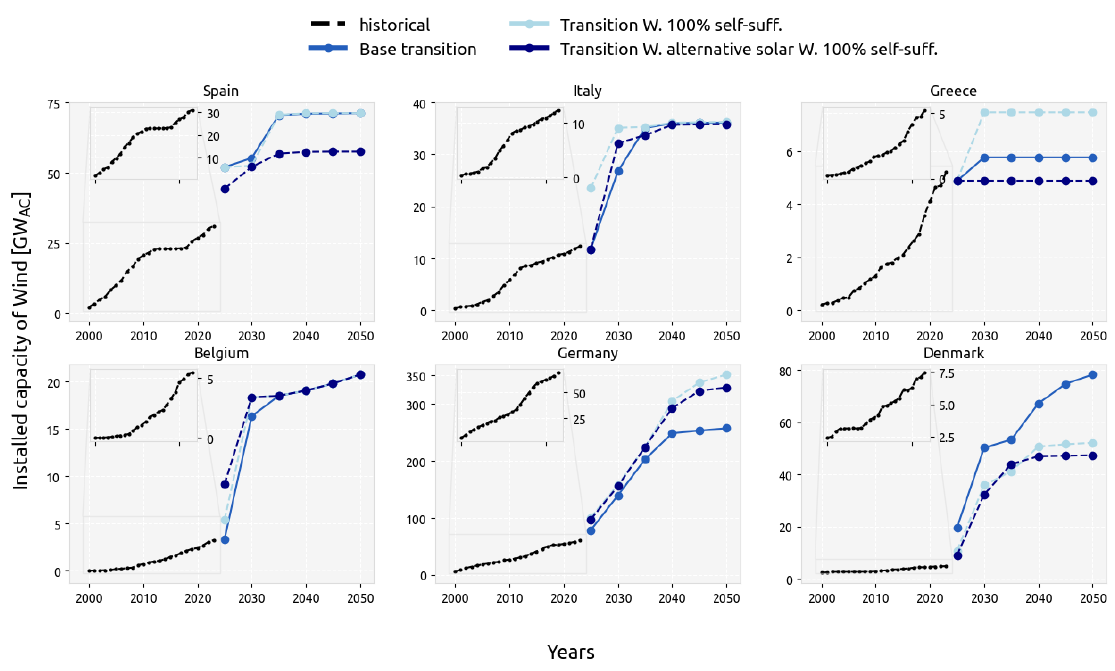}
   \caption{\textbf{} Historical (data from IRENA\cite{IRENA_2022}) and future cumulative installed capacity of onshore and offshore wind for selected European countries modeled under base transition, transition with a 100\% self-sufficiency target, and transition with selected alternative solar configurations under a 100\% self-sufficiency target. Refer to Supplementary Figs. S24 and S25 for other countries and the required installation rates.}
   \label{fig:6}
\end{figure*}

In light of historical data, the required cumulative PV capacity and annual installation rates for solar PV and wind technologies is very challenging for most countries. Nevertheless, the acceleration of installation rates in recent years is closing the gap between what would be needed and what has been shown to be possible. In previous work, Cherp and co-authors \cite{cherp2021national} raised concerns regarding the feasibility of the required installation rates based on historical data. For solar PV, they found a maximum annual rate of 0.6\% (interquartile range of 0.4-0.9\%) of the total electricity supply. However, their historical analysis is based on fitting S-curves to data up to 2019, a period in which wind and solar installations were mostly driven by policy-support measurements such as feed-in tariffs, especially until 2015. Figure \ref{fig:5} (inset figures) indicates that, for several European countries, after the end-of-policy-support stagnation, the historical data shows a second period in which capacity ramps up quickly pushed by technology competitiveness, reaching a maximum annual rate up to 3\% of annual electricity generation (see Supplementary Fig. \ref{fig:S26}). Still, under the self-sufficiency constraint many countries must reach installation rates up to five times higher than the maximum they have achieved so far to be able to meet their targets (see Supplementary Fig. \ref{fig:S23}). 

\begin{figure}[H]
\renewcommand*{\thefigure}{S\arabic{figure}} \renewcommand{\figurename}{Fig.} 
\includegraphics[width=1\textwidth, center]{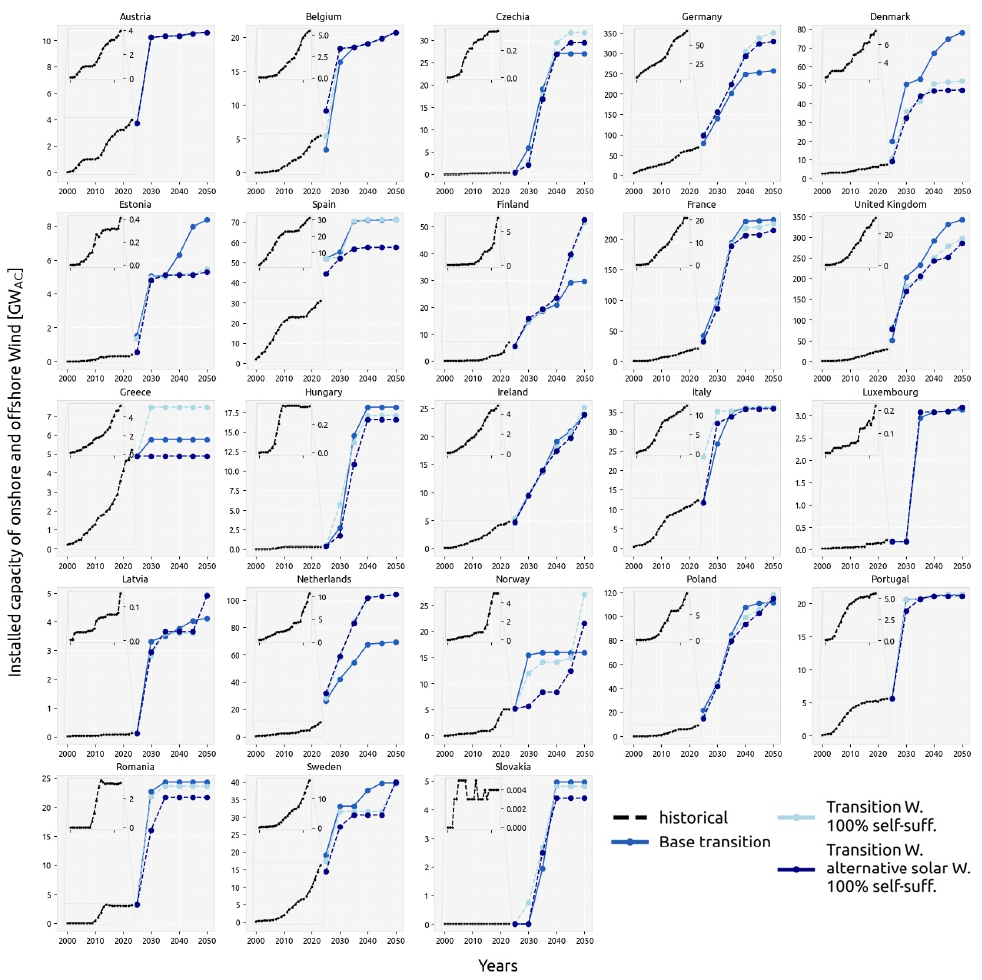}
\caption{Historical (data from IRENA \protect\citeS{IRENA_2022_S}) and future cumulative installed capacity of onshore and offshore wind for different European countries modeled under base transition, transition with a 100\% self-sufficiency target, and transition with selected alternative solar configurations under a 100\% self-sufficiency target.}
\label{fig:S24}
\end{figure}

The same comparison of historical and needed capacities for onshore and offshore wind (Fig. \ref{fig:6} and Supplementary Fig. \ref{fig:S24}) indicates that many European countries may be on a reasonable path for meeting their targets by 2050, but still higher ambitions are necessary for countries such as Germany and Belgium. We cannot see any dominant pattern regarding self-sufficiency targets and installation rates for wind, which can be attributed to wind resource and wind generation being non-uniformly distributed spatially compared to solar. However, many countries such as Spain, Germany, and Greece show a reduction in wind capacity when alternative solar configurations are available (Fig. \ref{fig:6}). Overall, most countries are in a better position to meet the required wind capacities than solar PV, although countries such as the UK and Denmark would still need a large ramp-up to happen in the near future.

For both wind and solar PV, growth rates might seem high compared to early historical rates, but costs are expected to continue decreasing in the future, which will enable fast growth \cite{bolinger2022levelized, wiser2021expert,way2022empirically}. Other factors that could accelerate growth are 
the rising social cost of carbon, which could in turn drive policies that increase carbon price in EU-ETS market \cite{tol2023social, rennert2022comprehensive, CO2_price_ETS}, and re-emergence of local cooperatives to gain public acceptance \cite{kirkegaard2023tackling, vuichard2022keep}. However, significant barriers remain, including grid congestion, slow permitting processes, and low social acceptance \cite{IEA2024_renewables}. Overcoming these obstacles through infrastructure upgrades, streamlined regulations, and community engagement will be essential for scaling up wind and solar PV installations.

\section{Conclusions and policy recommendations}

We investigate transition paths for European countries to achieve carbon neutrality by mid-century while simultaneously becoming self-sufficient. The findings show that overall system costs increase by just 2.1\% under a self-sufficiency constraint, but costs can rise up to 150\% for countries that were net-importers in an unconstrained scenario. Self-sufficiency also promotes a fairer land use distribution among countries, reducing potential social acceptance issues. Therefore, pursuing energy security alongside carbon neutrality in Europe could be achieved with limited total cost increases but significant national disparities.

Solar PV is projected to contribute the largest share of electricity generation (57\%), with further capacity expansions necessary under self-sufficiency due to its low cost and widespread resource availability. From a system perspective, horizontal single-axis tracking (HSAT) is favored as it extends solar generation hours, making it suitable for countries aiming for self-sufficiency. Additionally, using lower inverter capacities than the DC capacity of solar panels is cost-effective, with DC/AC ratios of 1.7 and 1.5 commonly chosen for Central and Southern Europe. The findings suggest revisions in macro-energy models, for which we recommend: (i) the inclusion of HSAT, (ii) proper representation of inverter sizing, and (iii) separate modeling for distributed PV systems versus utility-scale plants \cite{rahdan2024distributed}. In contrast, non-optimal solar orientations or delta configuration can be excluded since their value regarding displacing electricity generation in time diminishes when the system includes batteries.

Comparing the required future capacity of wind and solar for different countries to their historical values, we see that many countries will need a significant acceleration to reach a European-wide net-zero emissions target by 2050. This suggests that policymakers should place emphasis on avoiding potential barriers that slow down installation rates (e.g. inefficient regulatory frameworks or permitting processes, and  misuse of public consultation processes). The requirements for accelerating growth are lessened under the self-sufficiency target for previously net-exporter countries and increased for net-importer countries. Ultimately, while the final transition path chosen by each country will depend on many social and political factors, the results presented here emphasize that the goals of one country could have significant implications for many others, necessitating a common planning strategy for the European energy system. Each country should also develop plans not only for renewable technologies development, but synthetic fuel production, mainly hydrogen, oil, and methanol, as they are the main energy carriers that will be traded in the future.

\section{Methods}

We employ PyPSA-Eur, an open-source model, to simulate the sector-coupled European energy system. This model, described in detail in previous studies \cite{neumann2023potential, victoria_2022_S}, utilizes various datasets to represent demand across different sectors in Europe. It then constructs an energy system consisting of diverse generation technologies to meet this demand efficiently. The model optimizes the capacity and dispatch of all system elements, aiming to minimize total system costs while adhering to defined constraints. Covering the entire ENTSO-E area we use spatial resolution of 37 nodes with 370 regions for renewable potential estimation and capture a one-year period with 2-hour time steps. The model integrates electricity generation including solar, onshore and offshore wind, hydropower, nuclear, gas, and other fossil fuels, storage technologies such as batteries, and transmission grid as well as a simplified distribution grid. Additionally, it incorporates the heating, land transport, aviation, shipping, industry (including industrial feedstock), and agriculture sectors, considering their specific demands and incorporating relevant technologies like heat pumps, electric vehicles, and industrial processes (see Supplementary section S1).

PyPSA-Eur utilizes linear equations and constraints, resulting in a linear and therefore convex optimization problem. Two different modeling approaches are employed in this paper. Firstly, an overnight greenfield optimization imposing a 95\% CO\textsubscript{2} emissions reduction is used to identify which alternative PV configurations are more cost-effective. These are used in the second approach. Second, a myopic optimization is used to model the system from 2025 to 2050 with steps every 5 years. The initial year includes the current existing capacities of solar, wind, and nuclear in Europe. The CO\textsubscript{2} limit imposed in every planning horizon is determined based on a carbon budget for Europe corresponding to 1.7\degree C temperature increase and assuming exponential decay of CO\textsubscript{2} emissions (see Supplementary Fig. \ref{fig:S3}). All modeled scenarios are summarized in Table \ref{table:1}. 

\definecolor{Silver}{rgb}{0.776,0.776,0.776}
\begin{table}
\caption{Summary of scenario assumptions.}
\scriptsize
\centering
\begin{tblr}{
  width = 0.95\linewidth,
  colspec = {Q[95]Q[152]Q[148]},
  cell{1}{2} = {Silver},
  cell{1}{3} = {Silver},
  cell{3}{2} = {c=2}{0.7\linewidth},
  cell{4}{2} = {c=2}{0.7\linewidth},
  cell{7}{2} = {c=2}{0.7\linewidth},
  cell{8}{2} = {c=2}{0.7\linewidth},
  cell{9}{2} = {c=2}{0.7\linewidth},
  cell{10}{2} = {c=2}{0.7\linewidth},
  hline{1} = {2-3}{solid},
  hline{2,13} = {-}{solid},
  hline{3-12} = {-}{dotted}
}
 & \textbf{Solar Competition}                                      & {\textbf Transition to\\ \textbf self-sufficiency}                    \\
Optimisation                                & Overnight (greenfield)                                                  & Myopic (brownfield)                                                   \\
{Spatial\\resolution }                         & 37 nodes (370 regions for renewable)                                             &                                                                      \\
{Temporal\\resolution}                         & 2-hourly                                                                         &                                                                      \\
Technology costs                            & 2050                                                                             & Investment year                      \\
CO\textsubscript{2} target & 95\% reduction of emissions compared to 1990 level               & 1.7\degree C carbon budget with a net-zero goal by 2050 \\
Weather year                                & 2013                                                                             &                                                                      \\
Transmission expansion                      & Today’s capacity expandable by 10\%                              &                                                                      \\
Sectors                                     & {Electricity+heating+industry+agriculture+\\shipping/aviation/land transport)} &                                                                      \\
PV potential                                & 20 TW for utility PV and 1.5 TW for distributed PV                               &                                                                      \\
{Solar\\configurations }                       & {With all alternative solar configurations}                                        & With default/selected solar configurations                           \\ Self-sufficiency target   & 0\% / 100\%       & 0\% / 100\% target by 2050 (starting from 40\% in 2025)     
\end{tblr}
\label{table:1}
\end{table}

The optimization problem includes various constraints, which include limiting CO\textsubscript{2} emissions, limiting transmission expansion, and limiting CO\textsubscript{2} sequestration in underground stores, consisting of mostly salt caverns \cite{neumann2023potential}. Eq. (\ref{eq:1}) represents the energy balance constraint, ensuring equilibrium between demand and generation at every node \(i\) and time step \(t\).
\begin{equation*}
    \sum_r g_{i,r,t} + \sum_s \left(h_{i,s,t}^- - h_{i,s,t}^+ \right) + \sum_k \eta_{i,k,t} f_{k,t} + \sum_\ell K_{i\ell} f_{\ell,t} +
\end{equation*}
\begin{equation} \label{eq:1}
 \sum_p m_{p,t} = d_{i,t}   \quad \leftrightarrow \quad \lambda_{i,t} \quad \forall i,t,
\end{equation}

\begin{figure*}[t]
\renewcommand{\figurename}{Fig.} 
    \includegraphics[width=0.85\textwidth,center]{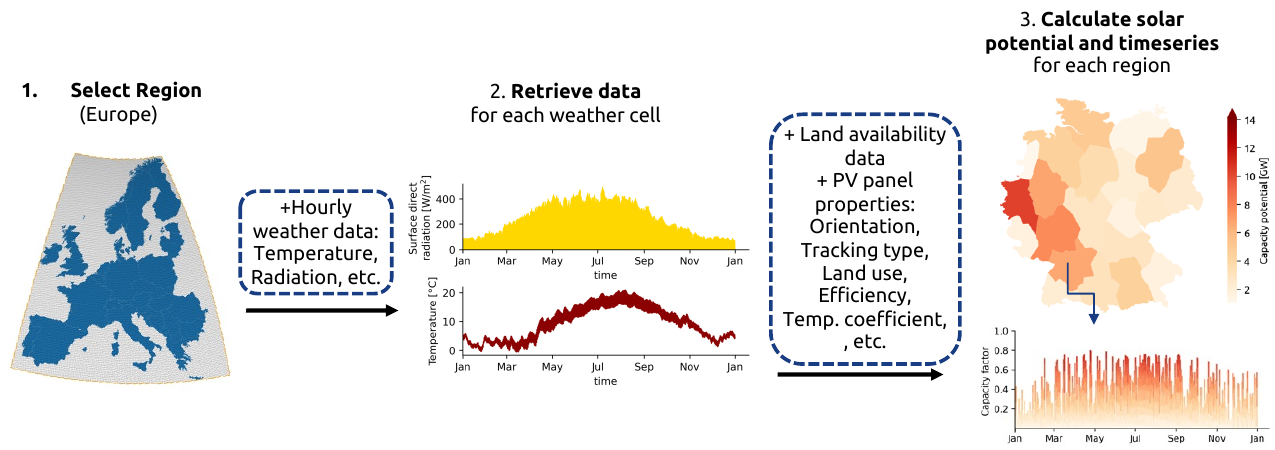}
   \caption{\textbf{} : Representation of the atlite package process for calculating the solar potential and the solar generation timeseries for different regions in Germany \cite{Hofmann2021}}
   \label{fig:7}
\end{figure*}

where \(g_{i,r,t}\) is generator dispatch of technology \(r\)  at time \(t\) and location \(i\), and \(h_{i,s,t}^-\) and \(h_{i,s,t}^+\) are the discharge and charge of storage unit \(s\), respectively. $K_{i\ell}$ is the incidence matrix of the energy transmission networks, such as AC and DC transmission lines or hydrogen pipelines, which has non-zero values equal to $-\eta_{i\ell}$/$\eta_{i\ell}$ when line ${\ell}$ is importing/exporting energy to or from node $i$, where $\eta_{i,\ell}$ is the efficiency of the  pipe or line, and $f_{\ell,t}$ is the imported/exported energy. \(f_{k,t}\) is dispatch of energy converter technology \(k\), such as heat pumps converting electricity to heat, and $\eta_{i,k,t}$ is the efficiency of the technology $k$ to represent conversion losses. \(m_{p,t}\) is the equivalent energy of fuel \(p\) such as methanol and oil that is imported to, or exported from, location \(i\). \(m_{p,t}\) is used to model energy carriers whose trade is assumed to be unlimited. \(d_{i,t}\) is demand from electricity, heating, transport, industry, and agriculture at location \(i\). $\lambda_{i,t}$ is the Lagrange multiplier of the constraint, which can be interpreted as the price of the respective energy carrier at location of bus \(i\) at time \(t\) \cite{neumann2023potential}. Following the approach proposed by van Greevenbroek et al.  \cite{vangreevenbroek2023enabling}, we implement the equity constraint, which requires that each country generates a share $c_{equity}$ of its own demand annually, as shown in Eq. (\ref{eq:2}). 
\begin{equation*}
\sum_{r,t}^{}g_{i,r,t}  +  \sum_{s,t} \left(h_{i,s,t}^- - h_{i,s,t}^+ \right)  +
\end{equation*}
\begin{equation} \label{eq:2}
\sum_{k,t} \eta_{i,k,t} \cdot f_{k,t}  \ge  c_{equity} \cdot \sum_{t} d_{i,t}   \quad\quad\quad  \forall  i  
 \in   nodes_{country}
\end{equation}

The constraint is individually applied to each country by totaling production and demand across all sectors for the entire year. As different energy forms (electricity, thermal energy, hydrogen, etc.) are not segregated, a country might generate electricity to offset its oil demand, which is discussed in the results section. However, implementing Eq. (\ref{eq:2}) as a constraint is complex due to various system losses, mainly from cyclic efficiencies of batteries, hydrogen stores, and water tanks. Eq. (\ref{eq:3}) modifies the constraint’s formulation from Eq. (\ref{eq:2}), by substituting demand from Eq. (\ref{eq:1}), to instead limit each country's net import relative to its total energy production, reducing complexity and computational burden.

\begin{equation*}
\left[ \sum_{r,t}^{}g_{i,r,t}+ \sum_{s,t} \left(h_{i,s,t}^- - h_{i,s,t}^+ \right) + \sum_{k,t} \eta_{i,k,t} \cdot f_{k,t} \right] \cdot (1-
\end{equation*}
\begin{equation} \label{eq:3}
\frac{1}{c_{equity}}) +\left[ \sum_{\ell,t} K_{i\ell} f_{\ell,t} + \sum_{p,t} m_{p,t} \right ] \leq 0  \quad  \forall i \in nodes_{country}
\end{equation}
Where the first bracket is the same as the left side of Eq .2 and represents energy production at node \(i\). The second bracket is the sum of net energy imports from gas, hydrogen, and electricity networks, plus the net imports of fuels such as oil and methanol, which are either used directly by industry, or by technology \(k\) for conversion of energy. It is noteworthy that as the constraint is implemented on an annual basis, there are still time periods when a country relies entirely on energy imports.

We use the atlite package \cite{Hofmann2021} to transform weather data into time series for onshore wind, offshore wind, and solar PV, as well as to estimate the potential capacity that can be installed in every region. The latter is estimated based on the available area considering suitable types of land (see Note S3.2 in Victoria et al. \cite{victoria_2022_S} for selected categories in Corine Land Cover database) and discounting the Natura 2000 protected areas. For onshore wind, 30\% of the available area is considered to estimate the potential and land-use of 10 MW/km\textsuperscript{2} is assumed. Wind velocity at 100 m is read from ERA5 reanalysis data \cite{ERA5} and scaled to hub height of the selected wind turbines (Vestas V112 3MW for onshore wind and NREL 5MW for offshore wind). The power curve of turbines is then used to calculate wind power generation \cite{bosch2018temporally, neumann2023potential}.

For solar PV, only 10\% of the available area is considered to estimate the potential and land-use of 102 MW/km\textsuperscript{2} \cite{DEA_2024} is assumed. Solar PV capacity factors are determined using satellite-aided  SARAH-2 data for irradiance and temperature and assuming a temperature-dependent module efficiency equal to 20\% at 25\degree C \cite{huld2010mapping}. For the default south-oriented PV configuration, the optimal inclination angle is selected for every country to ensure fair competition between fixed south-facing modules and other PV configurations. Analysis shows the optimal configuration with the highest generation for all countries is south-facing, with a tilt between 30\degree\ and 35\degree\ (see Supplementary Fig. \ref{fig:S6}). 

\begin{figure*}[t]
\renewcommand{\figurename}{Fig.} 
    \includegraphics[width=0.95\textwidth, center]{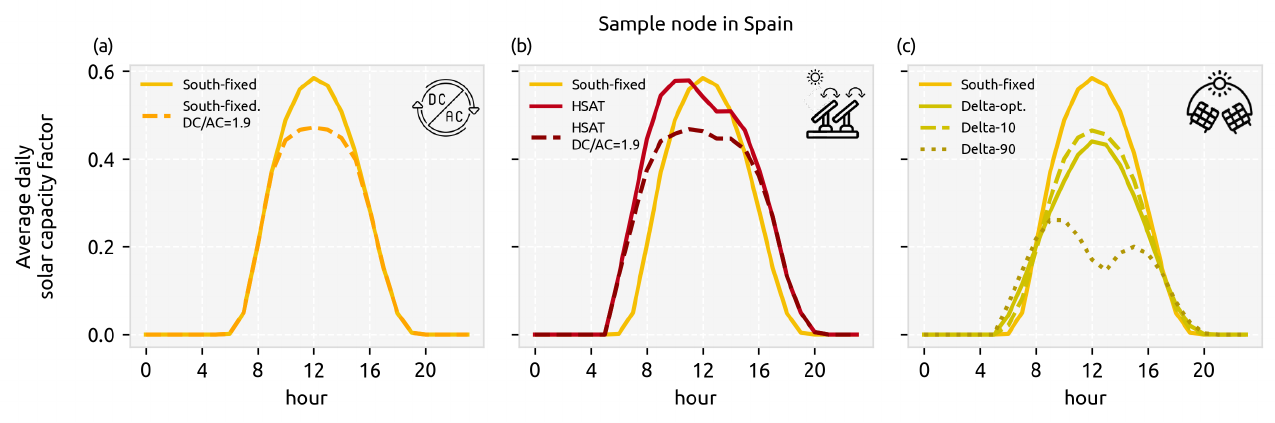}
   \caption{\textbf{} Hourly capacity factor for a day (average throughout the year) for a sample node in Spain for a) Fixed PV modules facing south and country-wise optimal inclination with 1 and 1.9 DC/AC ratio, b) Horizontal single-axis  tracking with 1 and 1.9 DC/AC ratio, and c) Delta configuration with 10\degree\/optimal/90\degree\ inclination. Note that optimal inclination refers to the country-wise optimised inclinations for south-facing PV modules.}
   \label{fig:8}
\end{figure*}
In our model, solar installations comprise utility PV power plants connected to high-voltage (HV) buses and distributed PV plants, also known as rooftop PV, connected to low-voltage (LV) electricity buses. Each region has an HV and LV bus connected by a link representing the distribution grid \cite{rahdan2024distributed}. On top of the south-oriented with optimal inclination, we analyze several alternative PV configurations. First, we consider inverter capacities lower than PV module capacity, known as inverter sizing. Second, we model horizontal single-axis tracking (HSAT) for utility PV plants. Third, we add delta configuration (with 10\degree\, optimal, and 90\degree\ tilt) for both utility and distributed PV. Fourth, we introduce east, west, southeast, and southwest-facing modules for rooftop PV. The daily generation profiles in Spain for a south-oriented configuration, HSAT, delta, and a south-oriented configuration with a DC/AC ratio of 1.9 are compared in Fig.\ref{fig:8}. Assumptions for these configurations are summarized in Table \ref{table:2}.

\begin{table}[htb!]
\caption{Alternative solar PV technologies.}
\scriptsize
\definecolor{Mercury}{rgb}{0.905,0.901,0.901}
\definecolor{Black}{rgb}{0,0,0}
\centering
\begin{tblr}{
  width = 0.95\linewidth,
  colspec = {Q[230]Q[175]Q[142]Q[132]Q[150]},
  row{1} = {Mercury},
  cell{7}{1} = {c=5}{0.95\linewidth},
  hlines,
  }
\textbf{Configuration}  & \textbf{{Investment} \tiny(\euro{}/kW\textsubscript{AC}) (2030) } & \textbf{O\&M {\tiny(\% of investment)}} & \textbf{Land use {\tiny ($\frac{MW_{AC}}{km^2}$)}}  & \textbf{Solar plant type}                               \\
South-oriented {(\tiny DC/AC=1)}    & 320.9 \cite{DEA_2024}                         & 2.47\% \cite{DEA_2024}        &  102\cite{DEA_2024}  & {Utility/ \\ Distributed PV}    \\
{South-oriented with inverter dimensioning {(\tiny DC/AC=1.25)} }    & 383.7  \cite{DEA_2024} & 2.47\% \cite{DEA_2024}   & 81.6\cite{DEA_2024}     & {Utility/ \\ Distributed PV} \\ {HSAT} {(\tiny DC/AC=1)}  & 377.5 \cite{DEA_2024}           & 2.28\% \cite{DEA_2024}  & 88.8\cite{NREL_land_use}                     & Utility PV                                  \\
{HSAT with \\ inverter \\ dimensioning \\ {(\tiny DC/AC=1.25)} }    & 454.5  \cite{DEA_2024} & 2.28\% \cite{DEA_2024}   & 71 \cite{NREL_land_use}     & {Utility/ \\ Distributed PV}  \\ {Delta \\ configuration}   & 250.3 \textsuperscript{†} & 2.47\% \cite{DEA_2024} & 138.9 \textsuperscript{†}   &  {Utility/ \\ Distributed PV} \\ {Lifetime of 40 years is assumed for PV modules and 10 years for inverters.\\Discount rate of 7\% is assumed for utility PV plants and 4\% for rooftop PV installations (see Supplementary Table S2 for cost assumptions of all years).\\ * The  estimates made here regarding land-use could vary greatly for different PV plants based on location, PV modules efficiency, etc. \cite{NREL_land_use, IRENA_2021}.\\
† Based on own calculations (see Supplementary section S3).} &  &    &    &   
\end{tblr}
\label{table:2}
\end{table}

\section{Acknowledgments}
P.R. and M.V. are partially funded by the AURORA project supported from the European Union’s Horizon 2020 research and innovation programme under grant agreement No. 101036418.

\subsection{Declaration of interests}
The authors declare no competing interests.

\subsection{Author contributions}

Conceptualization: P.R. and M.V.; Software: P.R., E.Z., and M.V.; Investigation: P.R. and M.V.; Methodology: P.R. and M.V.; Project Supervision: M.V.; Visualization: P.R.; Writing - original draft: P.R.; Writing - review and editing: P.R., M.V, E.Z..

\subsection*{Data and code availability}

The model is implemented by the open energy modeling framework PyPSA and makes use of the model PyPSA-Eur  v0.9.0 \cite{pypsa_docs} and the costs and technology assumptions included in the technology-data v0.4.0 \cite{pypsa_costs}. Scripts to reproduce the results and figures included in this paper are publicly available at: \href{/https://github.com/Parisra/Solar-Transition-Paper}{GitHub/Solar-Transition-Paper}.


\newpage
\addcontentsline{toc}{section}{References}
\bibliography{Bibliography.bib}

\clearpage
\onecolumn

\newgeometry{top=25mm ,
         bottom=25mm,
	left=15mm,
	right=15mm,
}

\addcontentsline{toc}{section}{Supplemental Information}
\beginsupplement
\clearpage

\section*{Supplemental Information}

\subsection*{S1. PyPSA-Eur model and self-sufficiency constraint}

We briefly go over some of the general features of the open-source PyPSA-Eur model here as detailed information regarding the optimization, the constraints, data sources, calculation of different demands, and assumptions made for modeling various technologies is already available in previous studies \protect\citeS{neumann2023_S,rahdan2024distributed_S} as well as the model documentation \citeS{pypsa_docs_S}. The main objective function for the optimization is to minimise the total annualized system costs, as shown in Eq. (\ref{eq:S1}).

\renewcommand{\theequation}{S\arabic{equation}}
\begin{equation*}
\min_{G,F,E,P,g,f} = \left[ \sum_{i,r}^{}c_{i,r}.G_{i,r} + \sum_{k}^{}c_{k}.F_{k}+ \sum_{i,s}^{}c_{i,s}.E_{i,s}+ 
\sum_{l}^{}c_{l}.P_{l}+ 
 \right.
\end{equation*}
\begin{equation} \label{eq:S1}
\left.\sum_{i,r,t}^{}w_{t}\left(\sum_{i,r}^{}o_{i,r}.g_{i,r,t}+\sum_{k}^{}o_{k}.f_{k,t}  \right)   \right]
\end{equation}
where \(c_{*}\) is capital cost of the component, \(o_{*}\) is operating cost of the component, \(G_{i,r}\) is generator capacity of technology \(r\) at location \(i\), \(E_{i,s}\) is energy capacity of storage \(s\) at location \(i\), \(P_{l}\) is transmission line capacity for line \(l\), \(F_{k}\) is power capacity of technology \(k\) for conversion and transportation of energy, \(g_{i,r,t}\) is generator dispatch of technology \(r\)  at time \(t\), and \(f_{k,t}\) is dispatch of technology \(k\) at time \(t\). Each time snapshot \(t\) is weighted by the time-step \(w_{t}\), and the sum of time-steps is one year. Costs for all technologies and the source for each data are available at the GitHub repository of PyPSA Technology Data \citeS{pypsa_costs_S}.

The assortment of linear constraints added to the optimisation problem is meant to represent different physical and societal limitations in the real-world energy system such as the maximum renewable potential in every region based on land availability, maximum transmission expansion limit based on social acceptance, available renewable and non-renewable resources depending on the weather, and maximum discharging and charging rates for each storage technology. These constraints vary from general to very specific, such as what percentage of the electric vehicles in the system can be used as batteries in each time-step. Other constraints in the model can help define a scenario's goals, such as the carbon emissions target or limiting the usage of gas in the system.

\begin{figure}[H]
\renewcommand*{\thefigure}{S\arabic{figure}} \renewcommand{\figurename}{Fig.} 
\includegraphics[width=0.8\textwidth,center]{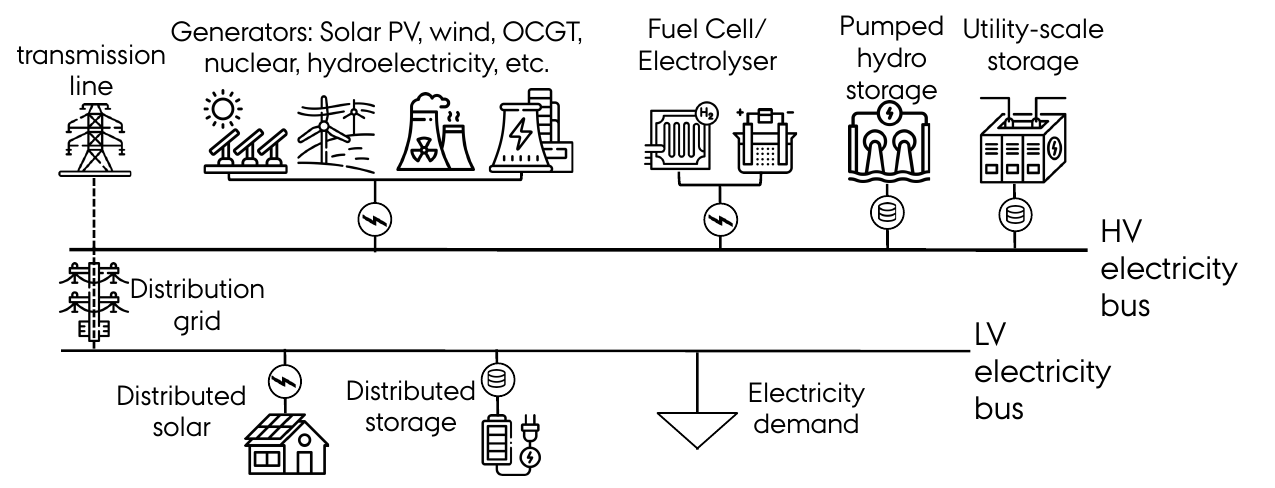}
\caption{Simplified representation of a single node when modeling the electricity sector. This figure has been designed using images made by Iconjam (utility battery), NeXore88 (Fuel Cell), and freepik (all other icons) from flaticon.com.}
\label{fig:S1}
\end{figure}

The self-sufficiency constraint, as shown in the main text, is meant to ensure that each country is able to produce a certain share of its own demand during the year. This constraint does not limit energy transmission, and when applied to the sector-coupled scenarios, does not differentiate between different energy carriers. To clarify how this constraint is implemented for various sectors, we can look at Fig. \ref{fig:S1}. This figure shows a simplified representation of how the electricity sector is modeled for each node. The local production of electricity is the sum of generation by the various technologies while system losses happen due to the cyclic efficiency of batteries, pumped hydro storage units, and hydrogen conversion technologies. Already we can see that including all these elements in a constraint is not a simple task. However, addition of other sectors and the interaction of energy carriers with each other complicates this even further.

\begin{figure}[H]
\renewcommand*{\thefigure}{S\arabic{figure}} \renewcommand{\figurename}{Fig.}  
\includegraphics[width=0.7\textwidth, center]{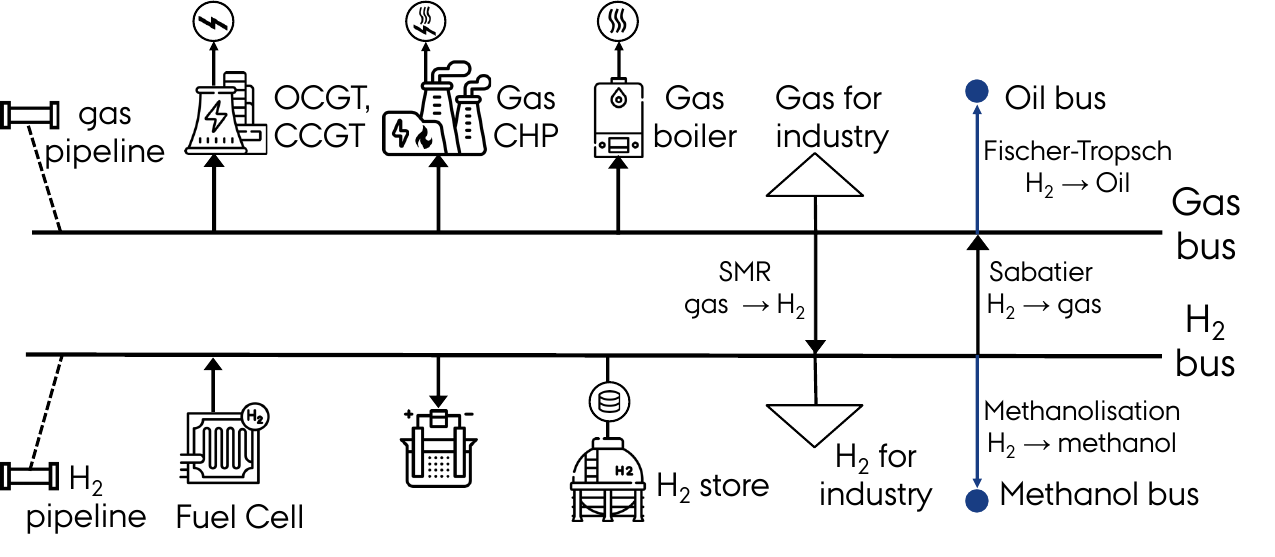}
\caption{Simplified representation of the gas and hydrogen bus of a single node when modeling all the sectors. For detailed information on different processes within the system refer to Neumann et al. \protect\citeS{neumann2023_S}. This figure has been designed using images made by NeXore88 (Fuel Cell) and freepik (all other icons) from flaticon.com.}
\label{fig:S2}
\end{figure}

Figure \ref{fig:S2} shows how the gas bus and the hydrogen bus in each node are connected to electricity and heat generators, storage technologies, pipelines to other nodes, separate demands such as gas for industry, and finally to each other in the form of chemical processes. Calculating all the losses for such a system is immensely difficult and will have to include conversion efficiencies from not just the gas and hydrogen technologies, but also ones for other carriers including oil, solid biomass, and methanol. Therefore, the constraint is modified as shown in the main text to instead limit the net imports of each country. It stands to reason that if a country is obliged to produce energy equal to its own demand during the year, the net imports of it would be lesser than or equal to zero. Using this method when implementing the constraint for Fig. \ref{fig:S2}, we again consider generators including open-cycle and closed-cycle gas turbines (OCGT and CCGT), and gas boilers for local production of energy, and calculate the net imports into the node using the hydrogen and gas pipelines. The production of fuel with chemical processes such as Fischer-Tropsch and methanolisation is also considered as local energy production in the constraint. 

For the main model, the forms of energy transport also include electricity transmission, biomass transport, and oil or methanol exchange. Similar to pipelines and transmission network, biomass transport includes a cost in our model, assuming to be done by trucks, but there is no upper limit to the capacity of biomass transport. Both methanol and oil exchange between countries are assumed to be done with negligible costs and without any limitations. It should be noted that this is still a computationally heavy constraint, and solving an optimisation problem with high spatio-temporal resolution under this constraint can take up to several hours and 100 GB of memory to complete.

\subsection*{S2. Carbon budget calculation}

\begin{figure}[H]
\renewcommand*{\thefigure}{S\arabic{figure}} \renewcommand{\figurename}{Fig.} 
\includegraphics[width=0.6\textwidth, center]{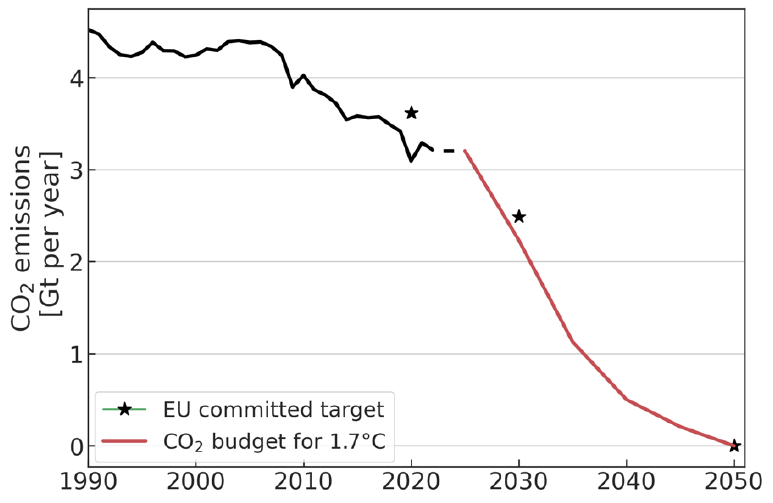}
\caption{Carbon budget assumed for Europe, equal to 29 Gt (+1.7\degree C). Carbon neutrality is required by 2050. \protect\citeS{neumann2023_S}}
\label{fig:S3}
\end{figure}

The carbon budget for Europe is calculated by taking the global carbon budget estimated to be left by 2025 for a temperature increase of 1.7\degree C, distributing it among countries by assuming equal per capita emissions, and including EU27 countries plus Norway, Switzerland, and the UK. The carbon budget left for Europe by 2020 is 45 Gt \citeS{zeyen2023endogenous_S}, and based on carbon emissions data for 2020-2022 \citeS{EEA_2024_S, UK_emissions_2024_S}, plus estimated carbon emissions for 2023 and 2024, the budget left for 2025 onwards is equal to 29 Gt. This budget is allocated to different time steps following an exponential decay with carbon neutrality imposed in 2050 (Fig. \ref{fig:S3})\citeS{victoria2020early_S}. Note that the carbon budget assumed only includes emissions from sectors that are included in the model, and the total emissions for 2023 and 2024 are predicted with a conservative assumption regarding the yearly decrease in emissions.

\subsection*{S3. Cost and land-use calculations for alternative solar configurations}

1. Inverter dimensioning

The capital cost, expressed in \({\scriptstyle (\frac{\text{\euro{}}}{MW_{AC}})}\), of a PV installation with an inverter DC/AC ratio equal to \(r\)  is calculated using Eq. (\ref{eq:S2}).
In a PV installation, one part of the capital cost depends on the DC capacity (e.g. PV modules, land-use costs, soft costs), as shown in Eq. (\ref{eq:S2}) as \({DC\ components}_{\frac{DC}{AC}=1}\), while the other part depends on the AC capacity of the plant (e.g. inverter, grid-connection), shown as \({AC\ components}_{\frac{DC}{AC}=1}\). Therefore, a PV power plant with DC/AC ratio of \(r\) will have costs equal to:

\begin{equation*}
{PV \ power \ plant \ capital cost}_{\frac{DC}{AC}=r} { (\textstyle \frac{\text{\euro{}}}{MW_{AC}})}= 
\end{equation*}
\begin{equation}  \label{eq:S2}
\textstyle {DC\ components}_{\frac{DC}{AC}=1} { (\frac{\text{\euro{}}}{MW_{DC}})\ }  . {\ r \ }  { (\frac{MW_{DC}}{MW_{AC}})} + {AC\ components}_{\frac{DC}{AC}=1} { (\frac{\text{\euro{}}}{MW_{AC}})}
\end{equation}

Costs for fixed panels and horizontal single-axis tracking (HSAT) with different inverter ratios as calculated by Eq. (\ref{eq:S2}) are shown in Table \ref{table:S1}. According to the Technology Data Catalogue \citeS{DEA_2024_S} published by the Danish Energy Agency (DEA), the share of AC components cost amounts to 19\% of total PV power plant capital cost in 2020. This percentage might increase in the future as PV modules are reducing their cost at a faster rate than inverters and other grid-related costs. The DEA estimates the share to be 23\% by 2050. 

2. Delta configuration

There is very limited information regarding the costs and land use of delta-shaped or east-west PV plants in literature \citeS{autarco_delta_S}. Therefore, we conduct our own calculations for these configurations. For land use, considering the higher packing factor for direct land use, and assuming direct land use accounts for 77\% of total land use for the plant \citeS{NREL_land_use_S},  30\% higher capacity density, relative to south-oriented systems, is assumed for the delta configuration.

As for costs, we follow a similar approach as the calculation done for inverter dimensioning. We assume each 1 kW of PV panels in the delta configuration is paired with a 0.66 kW inverter (this is the same as having a 1.5 DC/AC ratio). We estimate that the lower cost of the inverter and the lower land-use will result in a system that is 27\% cheaper than south-oriented systems, resulting in a capital cost equal to \( 250.3\  \text{\euro{}}/kW_{AC}\  (320.9\  \text{\euro{}}/kW_{AC} * 0.78 = 250.3\  \text{\euro{}}/kW_{AC})\) 

\begin{figure}[H]
\renewcommand*{\thefigure}{S\arabic{figure}} \renewcommand{\figurename}{Fig.} 
\includegraphics[width=0.9\textwidth,center]{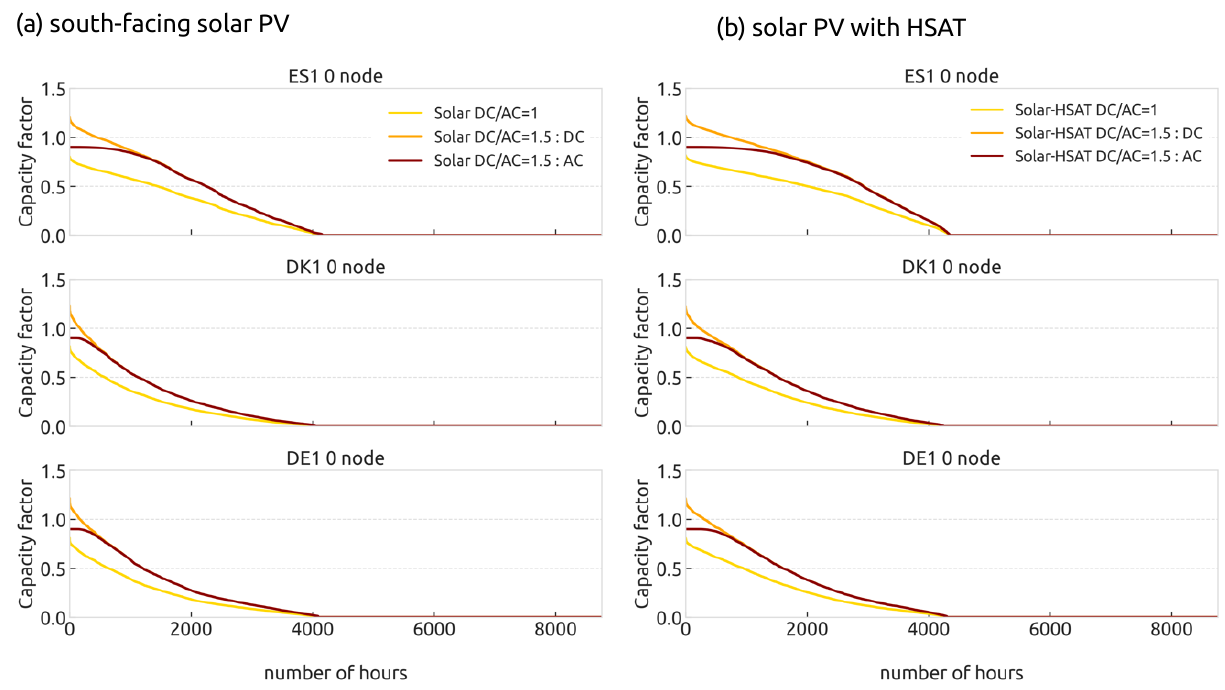}
\caption{Duration curve of capacity factors for a) south-facing fixed solar PV with DC/AC ratio of 1 and 1.5 (both DC and AC generation are shown), and b) solar PV with horizontal single-axis tracking (HSAT) with DC/AC ratio of 1 and 1.5 (both DC and AC generation are shown). }
\label{fig:S4}
\end{figure}

\begin{table}
\renewcommand*{\thetable}{S\arabic{table}}
\caption{Capital cost assumption for fixed panels and HSAT with different inverter ratios for 2025-2050.}
\definecolor{Silver}{rgb}{0.776,0.776,0.776}
{\scriptsize
\begin{tblr}{
  width = 0.9\linewidth,
  colspec = {Q[80]Q[135]Q[135]Q[73]Q[250]Q[110]},
  row{1} = {Silver},
  cell{2}{1} = {r=5}{},
  cell{2}{2} = {r=5}{},
  cell{2}{3} = {r=5}{},
  cell{7}{1} = {r=5}{},
  cell{7}{2} = {r=5}{},
  cell{7}{3} = {r=5}{},
  vline{2-6} = {1}{},
  hline{1-2,7,12} = {-}{},
  hline{3-6,8-11} = {4-6}{},
}
\textbf{Technology}       & \textbf{DC\\components cost {\tiny (\text{\euro{}}/kW\textsubscript{DC})} for 2025-2030-2040-2050} & \textbf{AC\\components cost {\tiny (\text{\euro{}}/kW\textsubscript{AC})} for 2025-2030-2040-2050} & \textbf{\textbf{DC/AC ratio}} & \textbf{Capital cost {\tiny (\text{\euro{}}/kW\textsubscript{AC})}\\ for 2025-2030-2040-2050 } & \textbf{Land use {\tiny (MW\textsubscript{AC}/km\textsuperscript{2})}}  \\
Fixed panels & 313.7 - 251.4 - 208.2 - 189.0 & 80.9 - 69.5 - 60.5 - 56.3                 
& 1  & 394.7 - 320.9 - 268.8 - 245.3  & 102   \\  
& & & 1.3  & 488.6 - 396.3 - 331.2 - 302.1  & 78.5 \\
& & & 1.5  & 551.2 - 446.6 - 372.9 -339.8  & 68  \\
& & & 1.7  & 613.8 - 496.8 - 414.5 - 377.6  & 60 \\
& & & 1.9  & 676.4 - 547.1 - 456.2  - 415.5 & 53.7 \\
HSAT~   & 377.2 - 307.9 - 258.9 - 236.9  & 80.9 - 69.5 - 60.5 - 56.3  & 
1  & 458.1 - 377.5 - 319.6 - 293.3   & 88.8  \\
& & & 1.3   & 571.1 - 469.9 - 397.3 - 364.4 & 68.3 \\
& & & 1.5   & 646.4 - 531.4 - 449.1 - 411.8 & 59.3  \\
& & & 1.7  & 721.8 - 593.0 - 500.8 - 459.2 & 52.2   \\
& & & 1.9   & 797.1 - 654.6 - 552.7 - 506.5   & 46.7        
\end{tblr} }
\label{table:S1}
\end{table}

\begin{figure}[H]
\renewcommand*{\thefigure}{S\arabic{figure}} \renewcommand{\figurename}{Fig.} 
\includegraphics[width=0.8\textwidth,center]{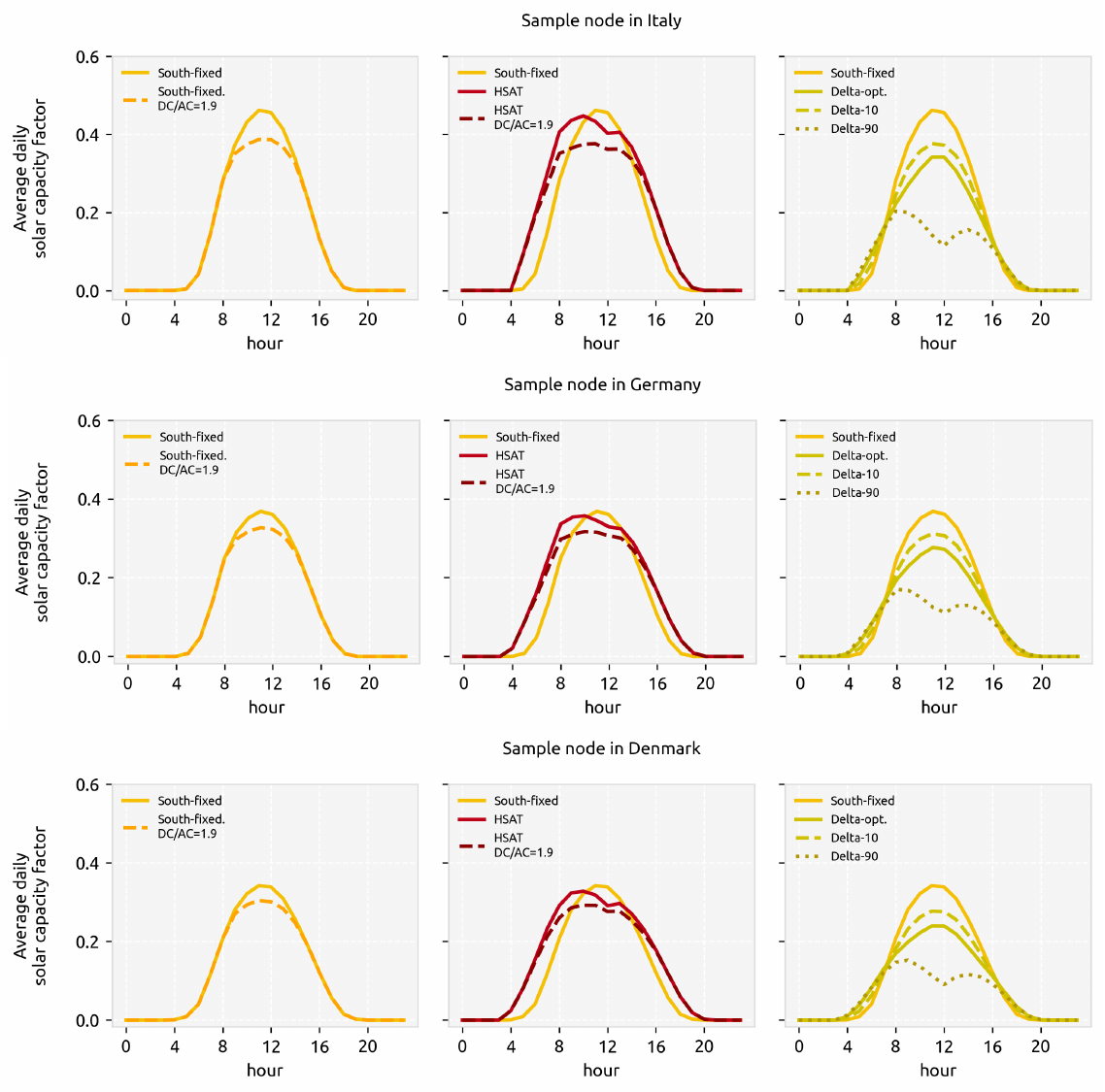}
\caption{Average daily capacity factors for sample nodes in Germany, Italy, and Denmark throughout the year for (left column) solar PV with fixed panels facing south and country-wise optimal slope with 1 and 1.9 inverter DC/AC ratio, (middle column) solar PV with horizontal single-axis tracking with 1 and 1.9 inverter DC/AC ratio, and (right column) delta configuration with 10\degree/optimal/90\degree inclination. The optimal inclination here refers to the best inclination for the south-facing panel that is optimised for each country.}
\label{fig:S5}
\end{figure}

\subsection*{S4. Solar generation for different panel configurations}

\begin{figure}[H]
\renewcommand*{\thefigure}{S\arabic{figure}} \renewcommand{\figurename}{Fig.} 
\includegraphics[width=0.8\textwidth,center]{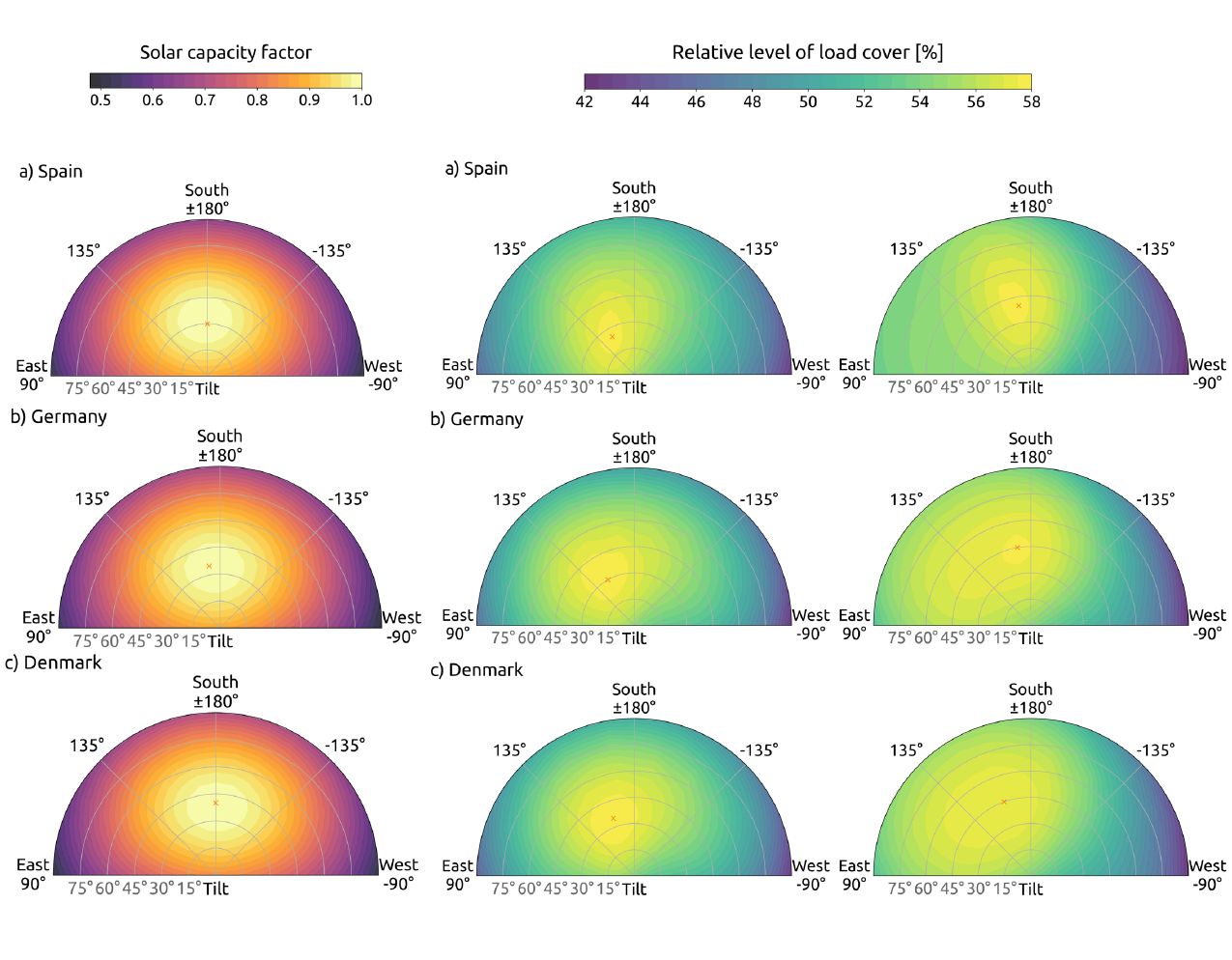}
\caption{(left) Annual capacity factor for different combinations of tilt and orientation angles for selected countries. Both radiation and the effect of temperature on panel efficiency are accounted for when calculating the capacity factor. The ratio of (middle) total demand and (right) sum of demand for peak 6-hour period that can be covered by solar generation with a fixed capacity under different combinations of tilt and orientation angles for selected countries. The fixed capacity is calculated by assuming the average demand is equal to the solar generation of a south-facing panel with 35\degree tilt. Figures were generated using SARAH-2 reanalysis data for year 2013.}
\label{fig:S6}
\end{figure}

\begin{figure}[H]
\renewcommand*{\thefigure}{S\arabic{figure}} \renewcommand{\figurename}{Fig.} 
\includegraphics[width=0.8\textwidth,center]{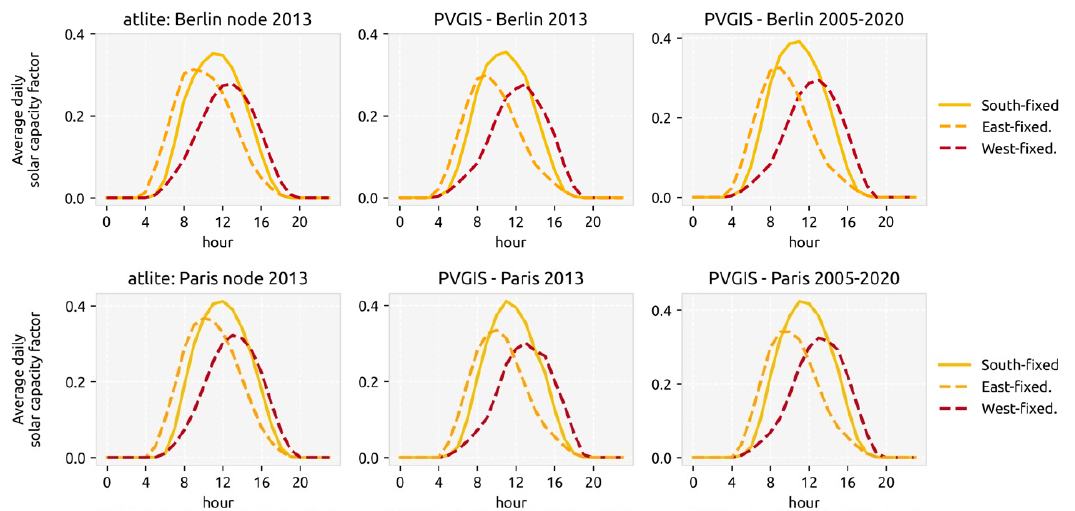}
\caption{Average daily capacity factor for the Berlin and Paris nodes, (left column) as calculated by the atlite package using radiation data from satellite-aided  SARAH-2 dataset and temperature data from ERA5 reanalysis dataset for weather year 2013, compared with daily average PV production data for Madrid and Berlin in (middle column) 2013 and (right column) 2005 to 2020 from the SARAH-2 dataset as calculated by PVGIS web interface. All figures indicate a higher PV production in the morning, but the bias from atlite is higher than the PVGIS data for 2013 \protect\citeS{huld2012new, pfeifroth2016validation, ERA5_S}. Note: the timestamp variation between ERA5 data and SARAH-2 dataset has been accounted for in the current study by shifting the SARAH-2 data by -30 minutes. For more details refer to PVGIS documentation on what the timestamps in each dataset represent \protect\citeS{PVGIS_documentation} . }
\label{fig:S7}
\end{figure}

We briefly look at the maximum available solar energy for panels with different inclinations and orientations. As seen from Fig. \ref{fig:S6}, the optimal orientation for a solar panel is always south, with an inclination between 30\degree and 35\degree. But looking at how much of the load can be covered by solar generation at every hour, east-facing panels are superior to west-facing ones. This is also true if we consider how much of the load can be covered by solar generation for the six hours with the highest electricity demand in each day. The higher load coverage for east-facing panels is due to higher solar generation in morning, as shown in Fig. \ref{fig:S7}. The higher average solar radiation in the morning in Europe is the main cause behind the superiority of east-facing panels (also seen in Fig. \ref{fig:2} of main text), which could be due to better average weather conditions. The non-symmetry of solar radiation is observed in other weather years as well. 

\subsection*{S5. Detailed cost, capacity, and import/export breakdown for all scenarios}

\begin{figure}[H]
\renewcommand*{\thefigure}{S\arabic{figure}} \renewcommand{\figurename}{Fig.} 
\includegraphics[width=0.8\textwidth, center]{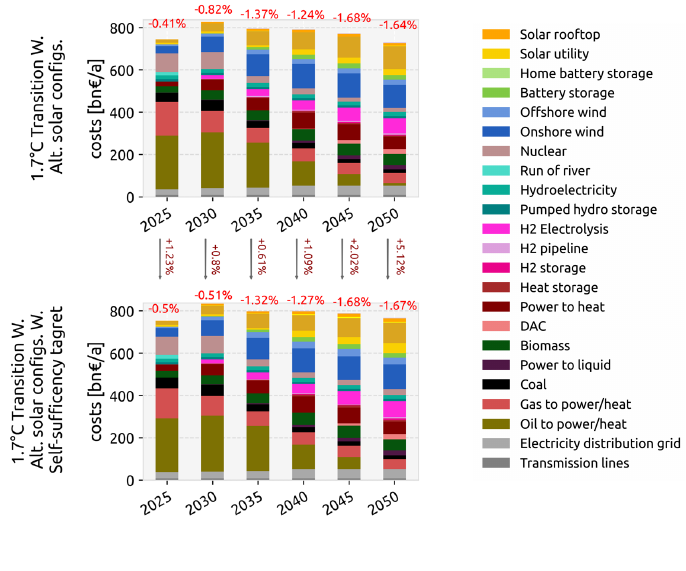}
\caption{Changes in total system costs during the transition under a 1.7\degree C temperature increase target with alternative solar configurations and with/without self-sufficiency target. The red numbers on each figure show the decrease in total system costs for scenarios with alternative solar configurations relative to scenarios without them, which are shown in Fig. \ref{fig:1} of the main text. For example, in year 2025, adding alternative solar configurations reduces total system cost for the ``1.7\degree C Transition with Self-sufficiency Target" scenario by 0.5\%. The total system cost is lowered by an average of 1.43\% when alternative solar configurations are added to the base transition, and by an average of 1.39\% when alternative solar configurations are added to the transition with self-sufficiency target. The discount rate used to calculate total cost throughout the whole transition is 7\%. Detailed information on the capital cost and lifetime assumed for each technology are available in the GitHub repository of PyPSA Technology Data \protect\citeS{pypsa_costs_S}.}
\label{fig:S8}
\end{figure}

\begin{figure}[H]
\renewcommand*{\thefigure}{S\arabic{figure}} \renewcommand{\figurename}{Fig.} 
\includegraphics[width=0.9\textwidth, center]{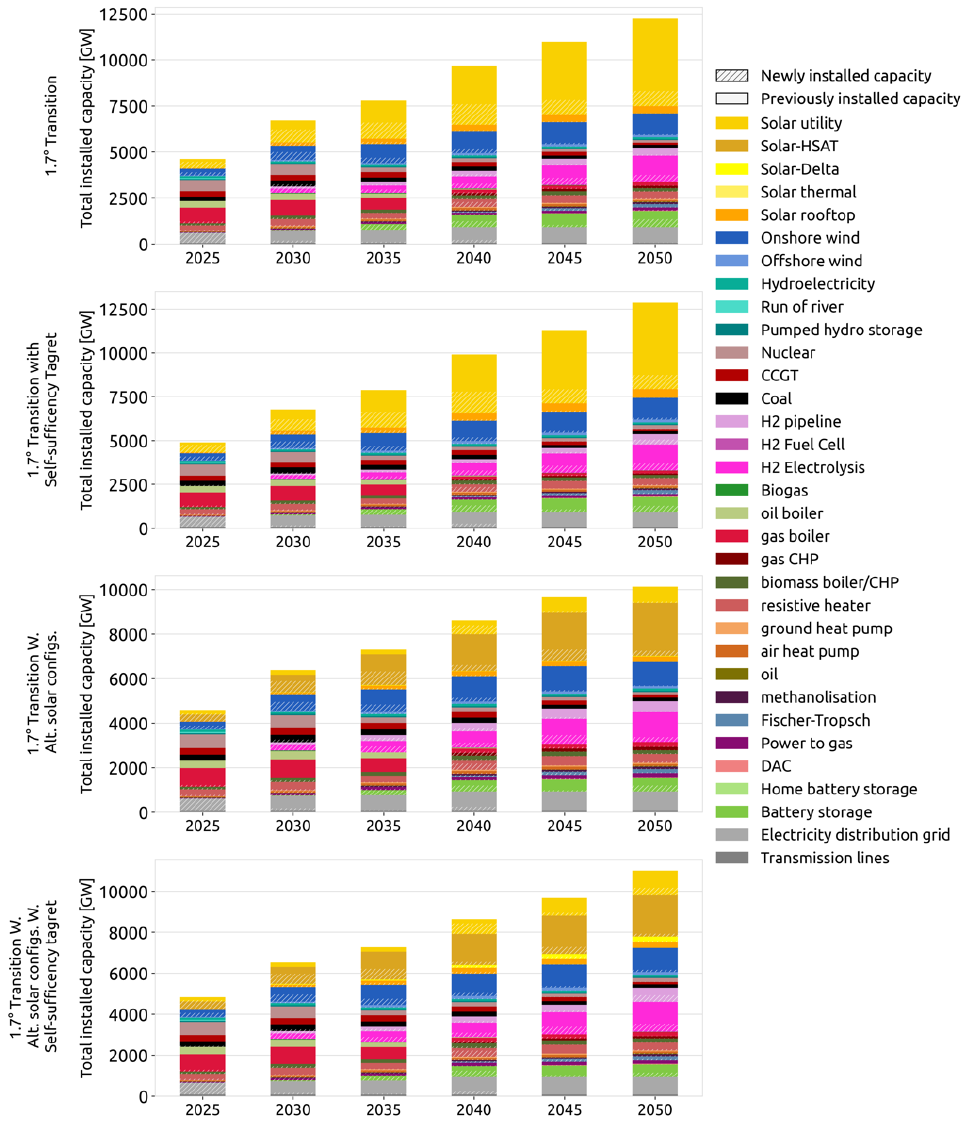}
\caption{Changes in Total installed capacity during the transition under a 1.7\degree temperature increase target for all scenarios. The installed capacity in each investment period is divided into previously installed capacity (brownfield) and newly installed capacity. The previously installed capacity at the beginning of the transition (2025) is taken from datasets provided by IRENA \protect\citeS{IRENA_2021_S}.}
\label{fig:S9}
\end{figure}

\begin{figure}[H]
\renewcommand*{\thefigure}{S\arabic{figure}} \renewcommand{\figurename}{Fig.} 
\includegraphics[width=0.8\textwidth, center]{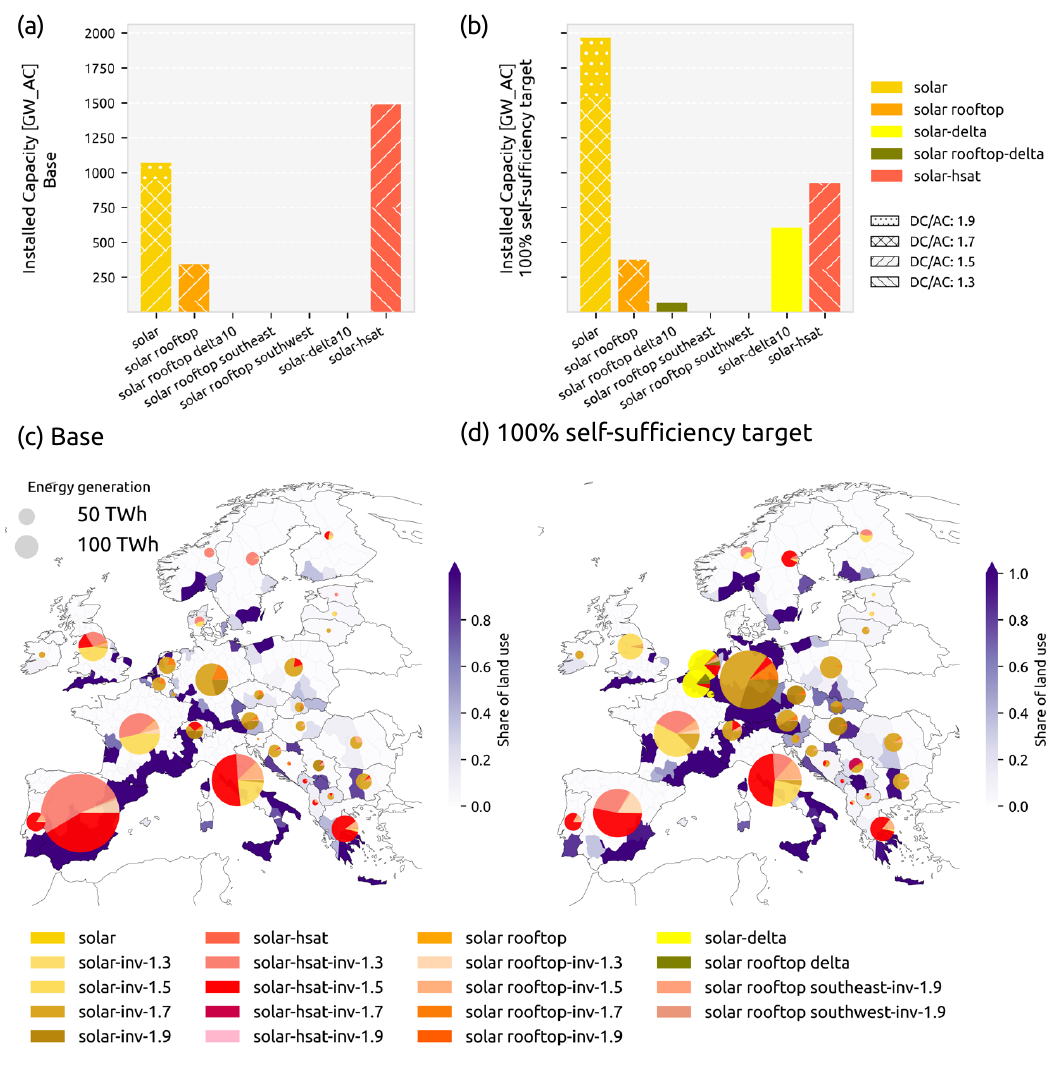}
\caption{Results for the overnight scenarios (95\% carbon emissions reduction, greenfield assumptions, technology costs for 2050) where the cost-efficiency of 19 alternative solar configurations is examined. The installed capacity of all alternative configurations for the overnight scenarios is shown at the top row a) without self-sufficiency target and b) with self-sufficiency target  }
\label{fig:S10}
\end{figure}

\begin{figure}[H]
\renewcommand*{\thefigure}{S\arabic{figure}} \renewcommand{\figurename}{Fig.} 
\includegraphics[width=0.8\textwidth, center]{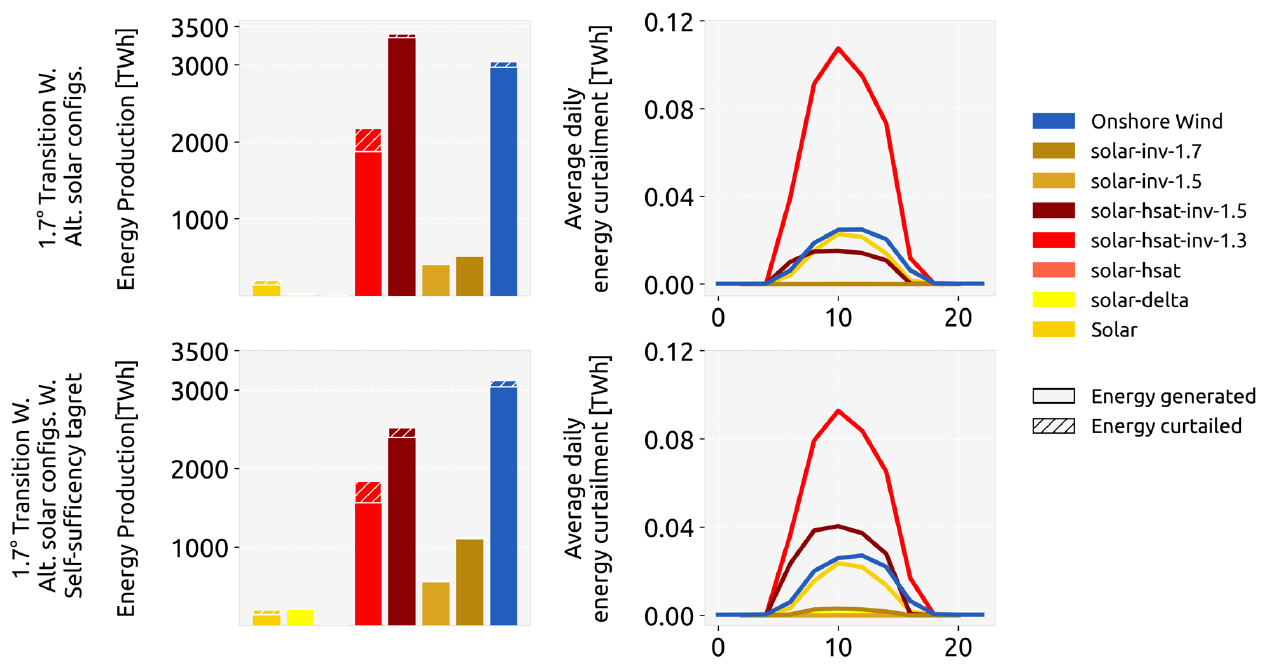}
\caption{Results for the transition scenarios showing a) Total energy generation from solar PV configurations and onshore wind plus the amount of curtailed energy, and b) average daily energy curtailment. The high curtailment ratio of HSAT with DC/AC ratio of 1.3 indicates the extra energy at noon is not deemed necessary by the system. This is the reason HSAT with DC/AC ratio of 1.5 has a higher installed capacity than HSAT with DC/AC ratio of 1.3 in many countries, even though DC/AC ratio of 1.3 is more cost-efficient overall (refer to Fig. \ref{fig:2} of main text). As discussed in the main text, total generation vs. capital cost is not the only deciding factor when it comes to the selection of a configuration, as evidenced by the results shown in Fig. \ref{fig:3} and Fig. \ref{fig:4} of the main text. }
\label{fig:S11}
\end{figure}

\begin{figure}[H]
\renewcommand*{\thefigure}{S\arabic{figure}} \renewcommand{\figurename}{Fig.} 
\includegraphics[width=1\textwidth, center]{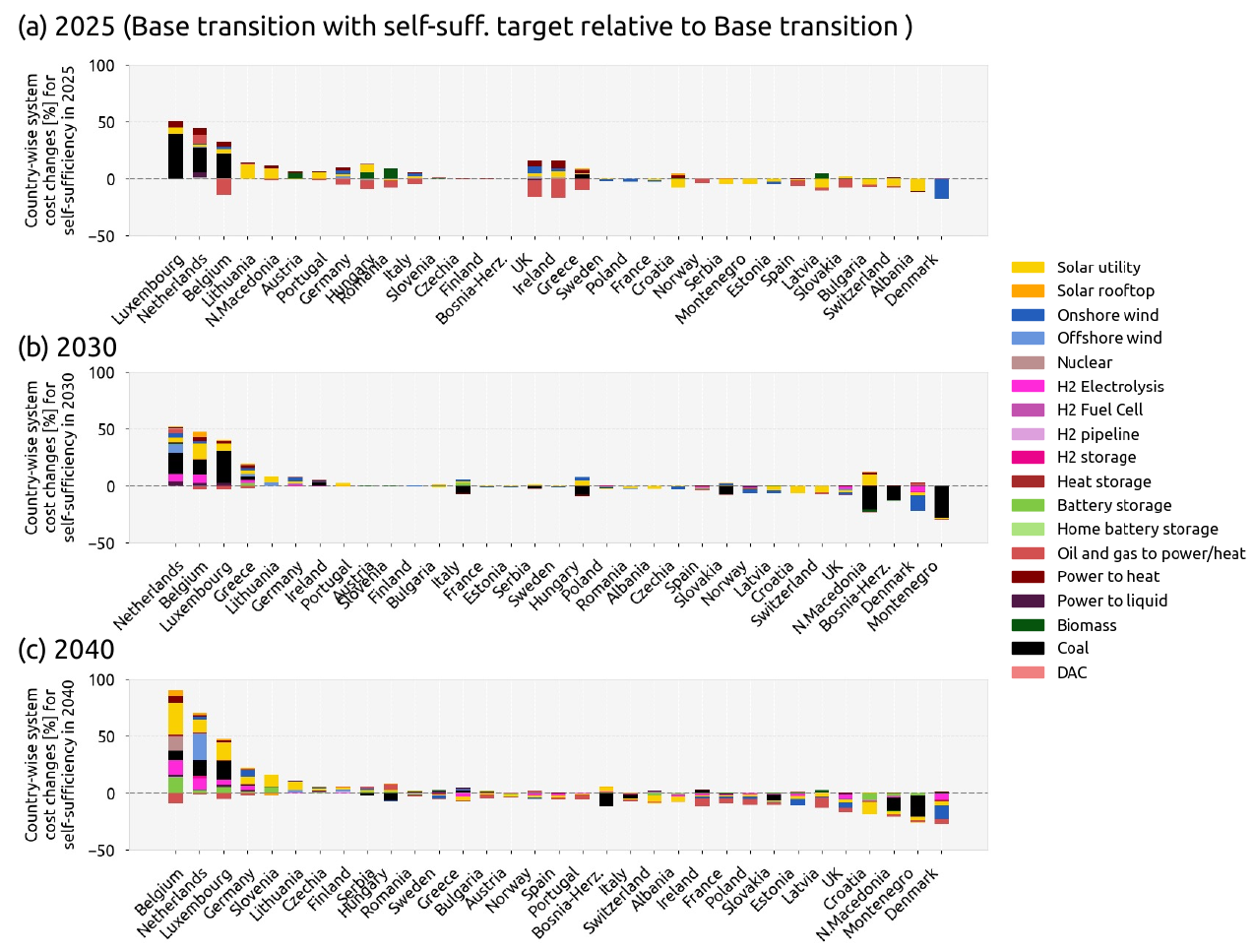}
\caption{Country-wise total system cost change for base transition with self-sufficiency target relative to the base transition. The increase/decrease in total country costs is shown as a percentage, and the share of each component from this increase is also shown for each country. Countries in each figure are ordered from one with the highest increase to the one with the highest decrease in costs. Transmission lines, H2 pipelines, gas pipelines, and other components where the costs would be shared between countries are not shown in the figure.}
\label{fig:S12}
\end{figure}

\begin{figure}[H]
\renewcommand*{\thefigure}{S\arabic{figure}} \renewcommand{\figurename}{Fig.} 
\includegraphics[width=1\textwidth, center]{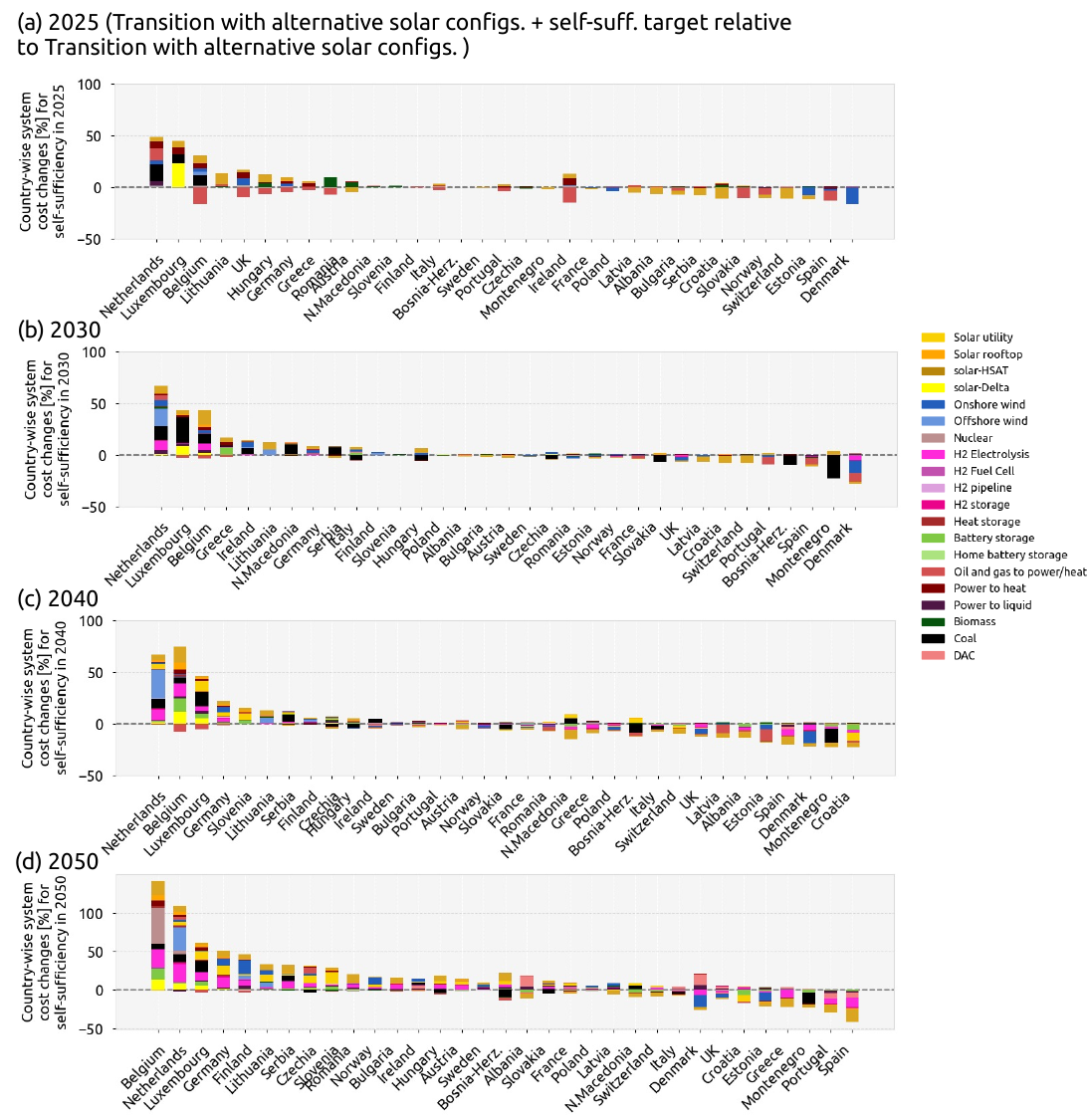}
\caption{Country-wise total system cost change for transition with alternative solar configurations and self-sufficiency target relative to the transition with self-sufficiency target. The increase/decrease in total country costs is shown as a percentage, and the share of each component from this increase is also shown for each country. Countries in each figure are ordered from one with highest increase to the one with the highest decrease in costs. Transmission lines, H2 pipelines, gas pipelines, and other components where the costs would be shared between countries are not shown in the figure.}
\label{fig:S13}
\end{figure}

\begin{figure}[H]
\renewcommand*{\thefigure}{S\arabic{figure}} \renewcommand{\figurename}{Fig.} 
\includegraphics[width=1\textwidth, center]{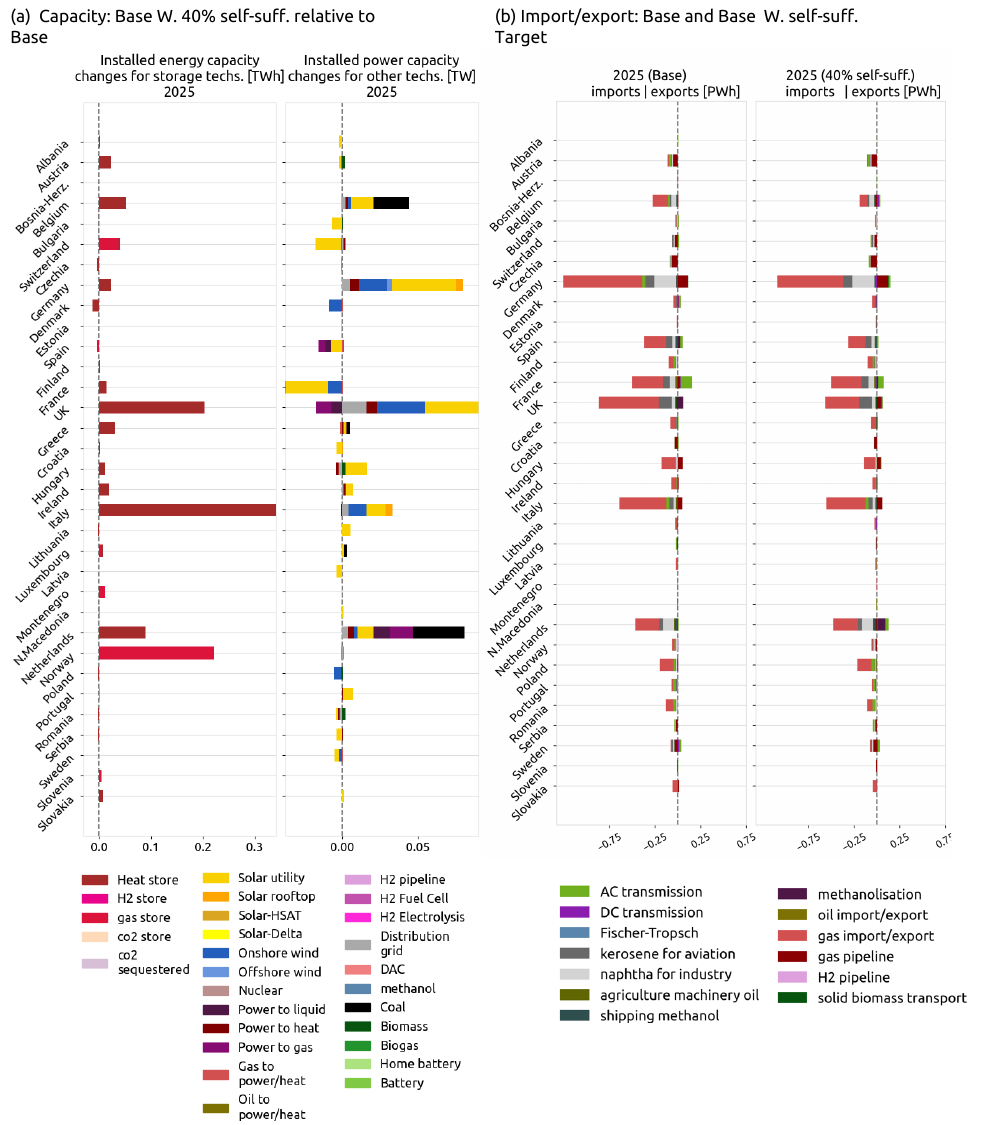}
\caption{Changes in the a) total installed capacity of different technologies for the base scenario with 40\% self-sufficiency target relative to the base scenario in 2025, and b) import and export of different countries for the base scenario and base scenario with 40\% self-sufficiency target in 2025. Power to gas includes capacities that produce hydrogen using steam methane reforming (SMR) and methane using the Sabatier process \protect\citeS{neumann2023_S}. The increase in the capacity of distribution grid under the self-sufficiency scenario is due to the higher electricity demand for heat pumps at the low-voltage bus due to the lower gas imports limiting the use of gas boilers to provide heating.}
\label{fig:S14}
\end{figure}

\begin{figure}[H]
\renewcommand*{\thefigure}{S\arabic{figure}} \renewcommand{\figurename}{Fig.} 
\includegraphics[width=1\textwidth, center]{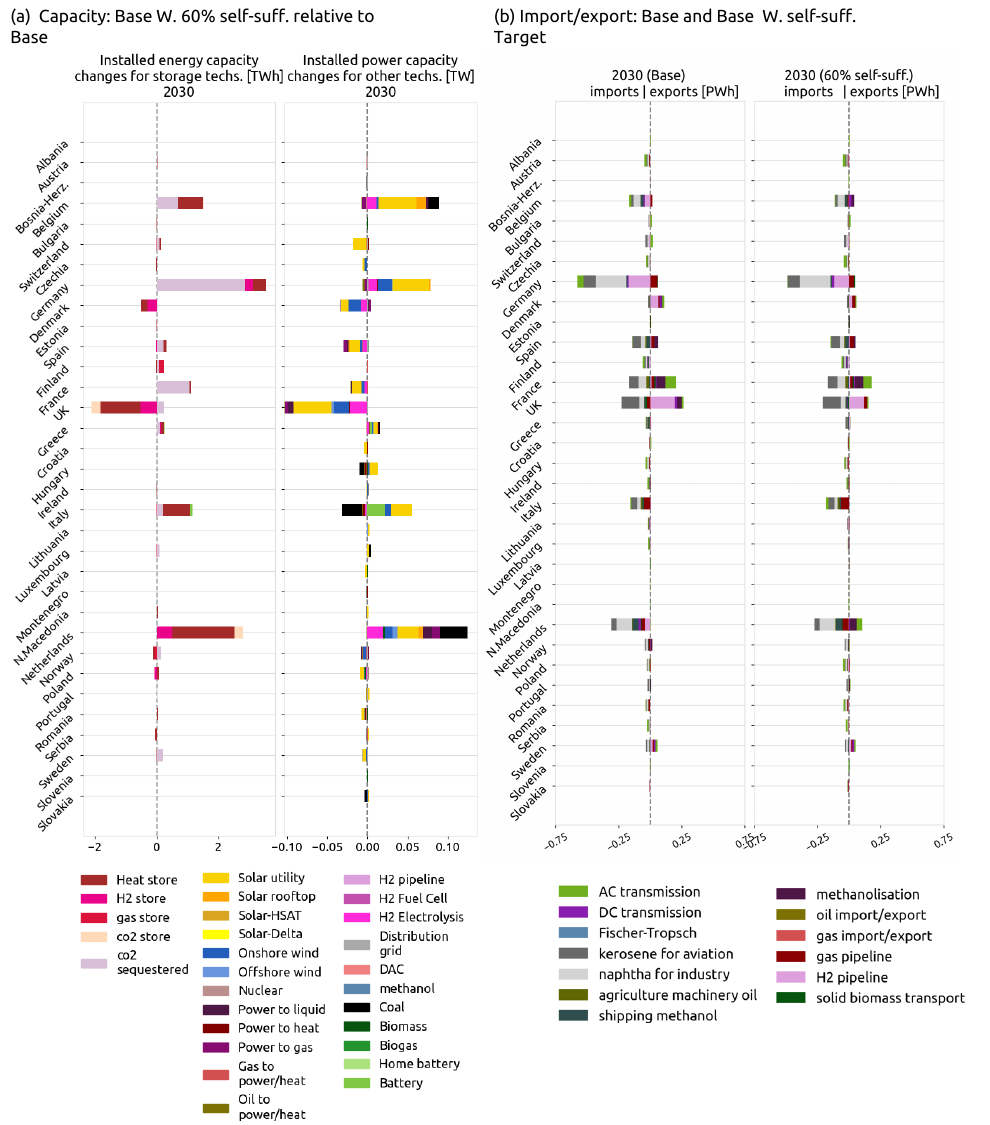}
\caption{Changes in the a) total installed capacity of different technologies for the base scenario with 60\% self-sufficiency target relative to the base scenario in 2030, and b) import and export of different countries for the base scenario and base scenario with 60\% self-sufficiency target in 2030.}
\label{fig:S15}
\end{figure}

\begin{figure}[H]
\renewcommand*{\thefigure}{S\arabic{figure}} \renewcommand{\figurename}{Fig.} 
\includegraphics[width=1\textwidth, center]{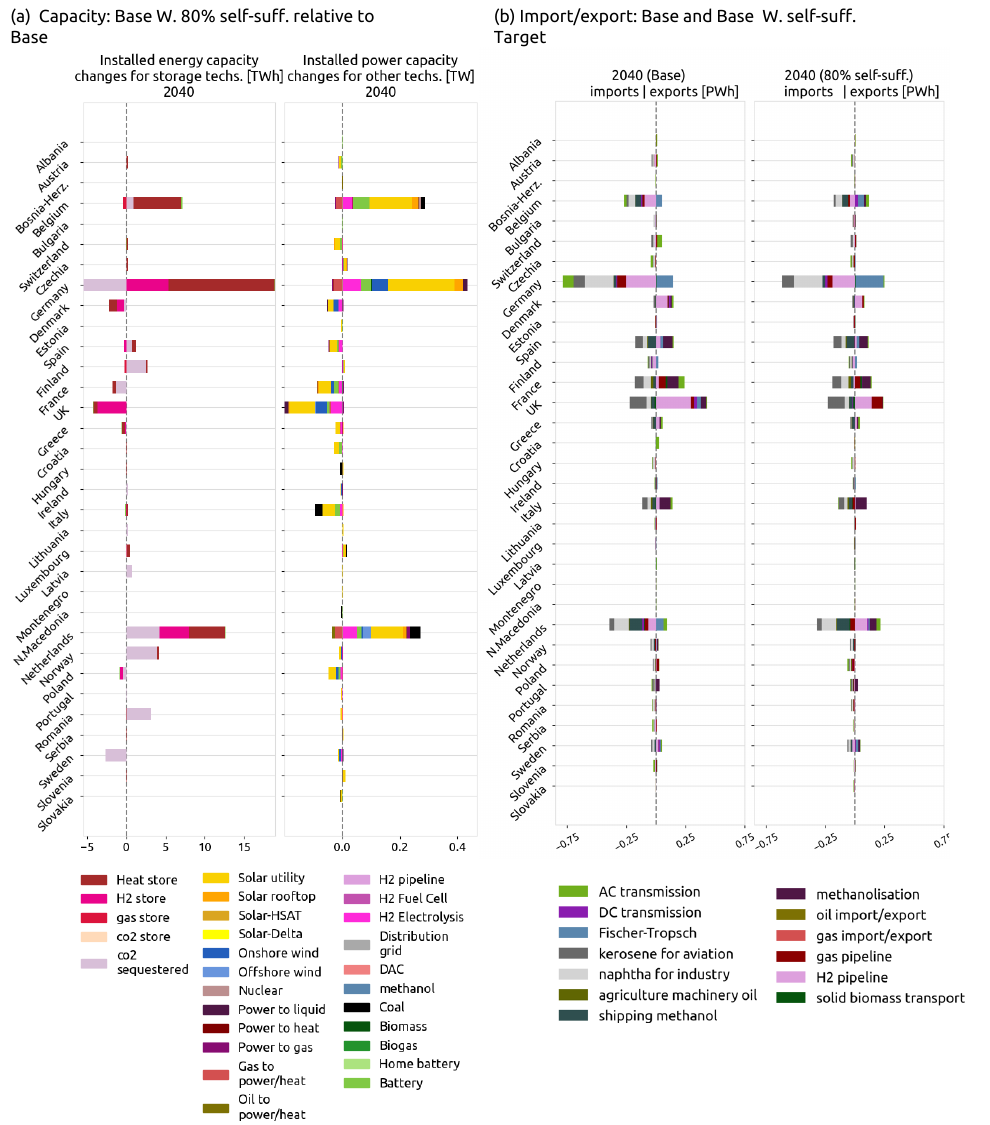}
\caption{Changes in the a) total installed capacity of different technologies for the base scenario with 80\% self-sufficiency target relative to the base scenario in 2040, and b) import and export of different countries for the base scenario and base scenario with 80\% self-sufficiency target in 2040.}
\label{fig:S16}
\end{figure}

\begin{figure}[H]
\renewcommand*{\thefigure}{S\arabic{figure}} \renewcommand{\figurename}{Fig.} 
\includegraphics[width=1\textwidth, center]{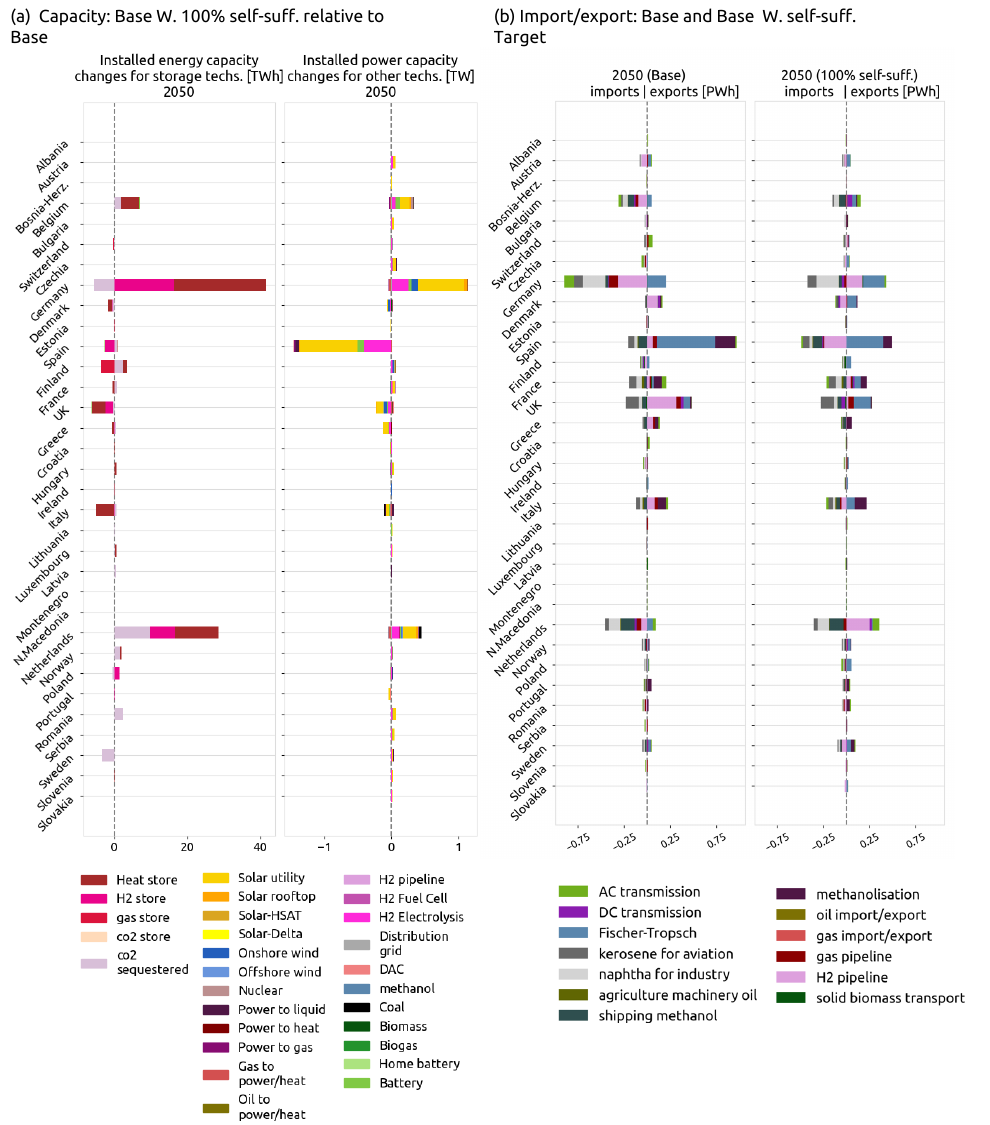}
\caption{Changes in the a) total installed capacity of different technologies for the base scenario with 100\% self-sufficiency target relative to the base scenario in 2050, and b) import and export of different countries for the base scenario and base scenario with 100\% self-sufficiency target in 2050.}
\label{fig:S17}
\end{figure}

\begin{figure}[H]
\renewcommand*{\thefigure}{S\arabic{figure}} \renewcommand{\figurename}{Fig.} 
\includegraphics[width=1\textwidth, center]{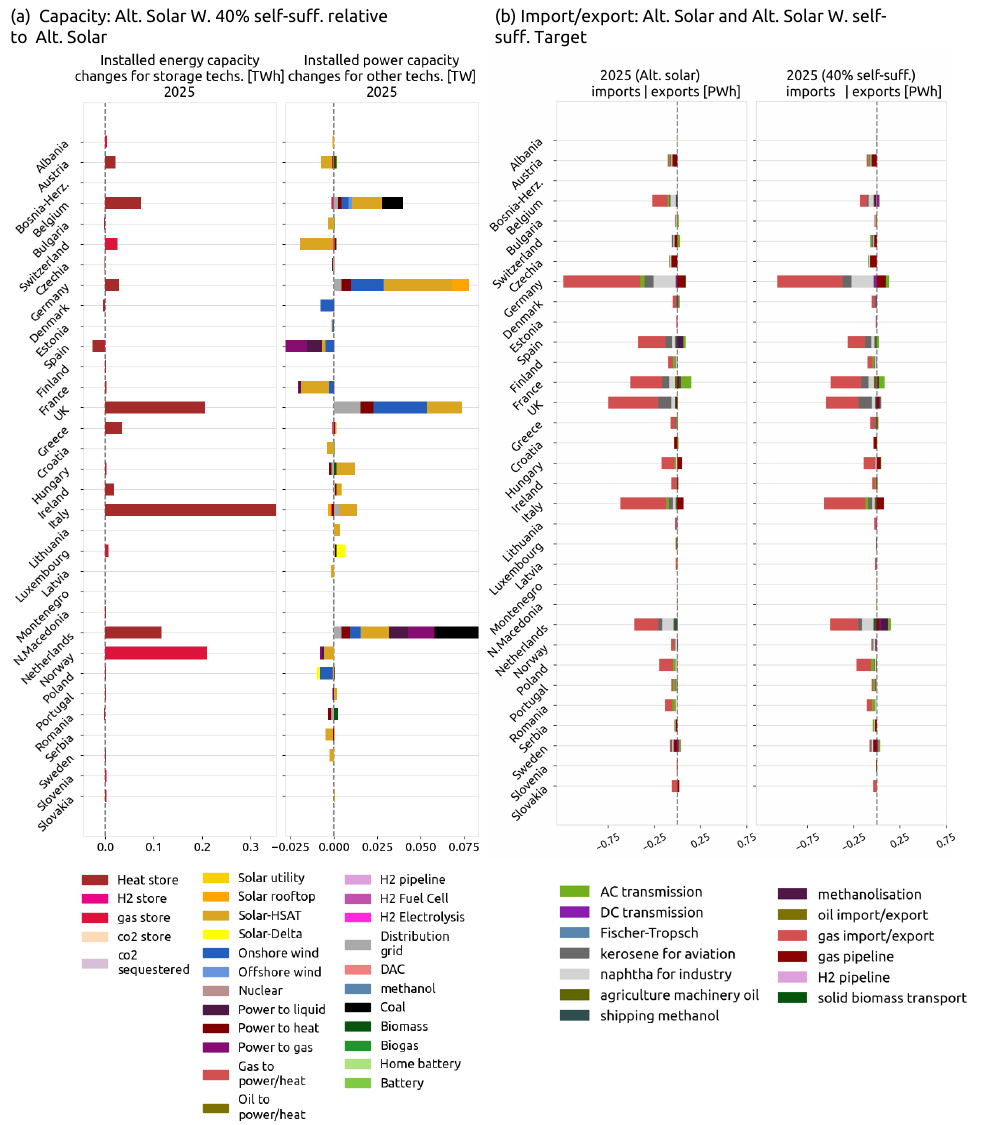}
\caption{Changes in the a) installed capacity of different technologies for the scenario with alternative solar configurations and 40\% self-sufficiency target relative to the scenario with alternative solar configurations in 2025, and b) import and export of different countries for the scenario with alternative solar configurations and the scenario with alternative solar configurations plus 40\% self-sufficiency target in 2025.}
\label{fig:S18}
\end{figure}

\begin{figure}[H]
\renewcommand*{\thefigure}{S\arabic{figure}} \renewcommand{\figurename}{Fig.} 
\includegraphics[width=1\textwidth, center]{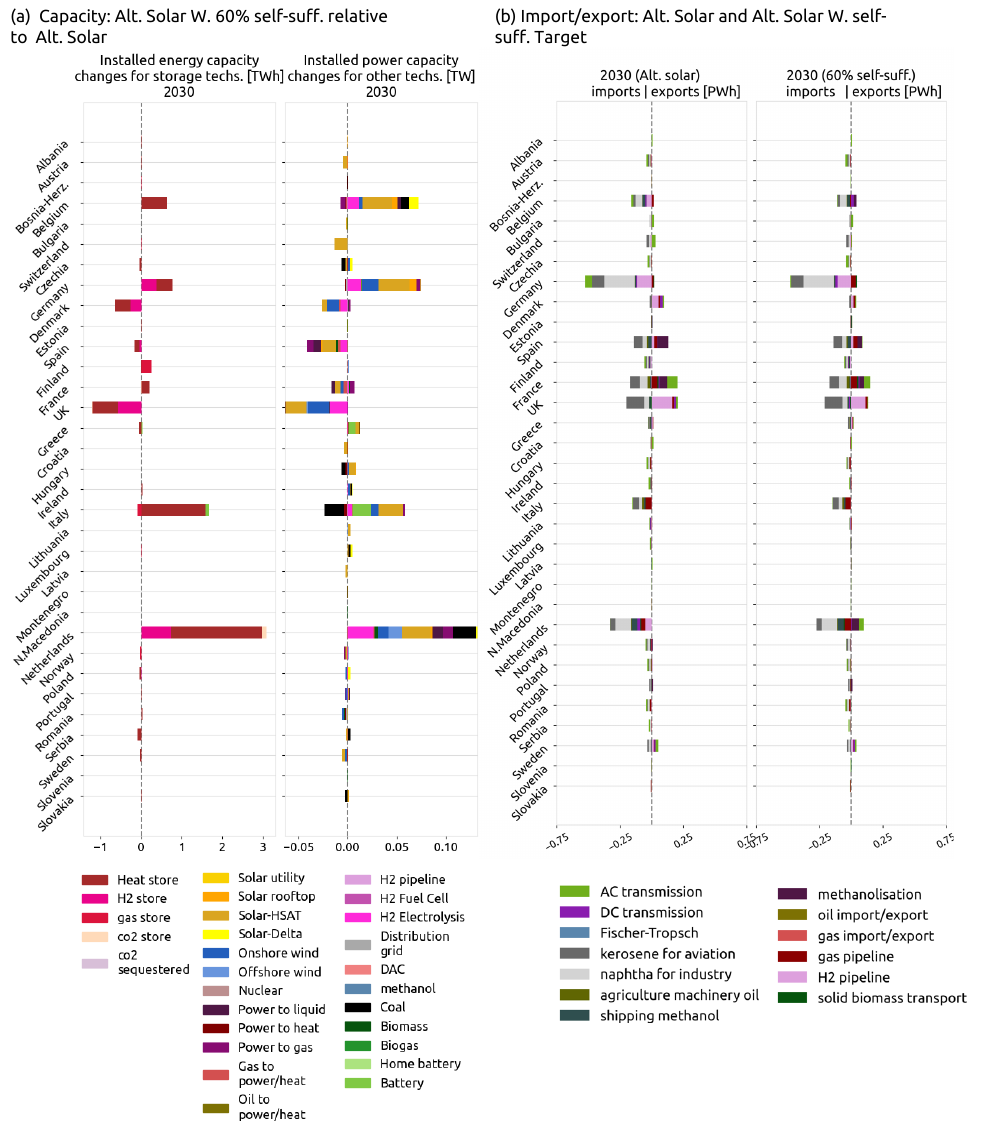}
\caption{Changes in the a) installed capacity of different technologies for the scenario with alternative solar configurations and 60\% self-sufficiency target relative to the scenario with alternative solar configurations in 2030, and b) import and export of different countries for the scenario with alternative solar configurations and the scenario with alternative solar configurations plus 60\% self-sufficiency target in 2030.}
\label{fig:S19}
\end{figure}

\begin{figure}[H]
\renewcommand*{\thefigure}{S\arabic{figure}} \renewcommand{\figurename}{Fig.} 
\includegraphics[width=1\textwidth, center]{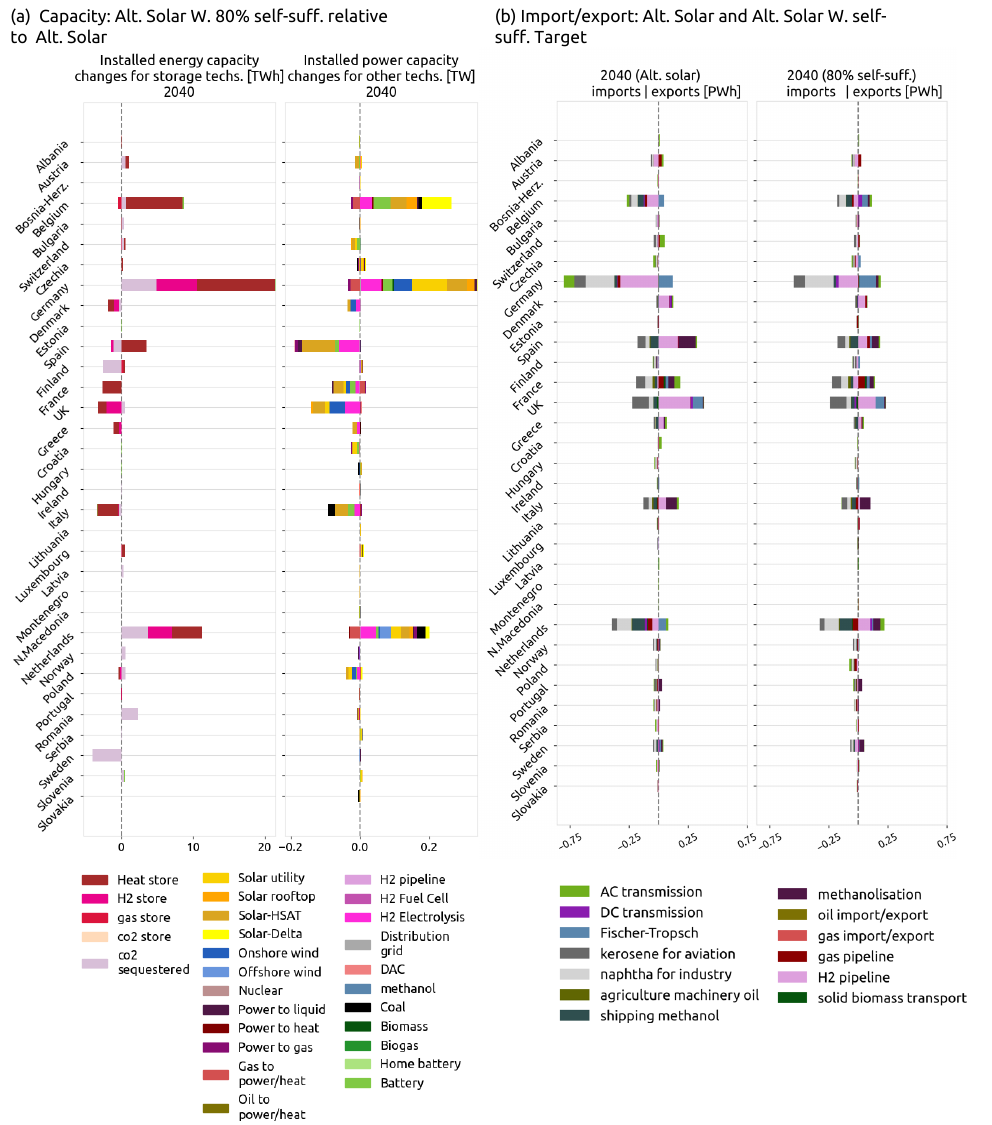}
\caption{Changes in the a) installed capacity of different technologies for the scenario with alternative solar configurations and 80\% self-sufficiency target relative to the scenario with alternative solar configurations in 2040, and b) import and export of different countries for the scenario with alternative solar configurations and the scenario with alternative solar configurations plus 80\% self-sufficiency target in 2040.}
\label{fig:S20}
\end{figure}

\begin{figure}[H]
\renewcommand*{\thefigure}{S\arabic{figure}} \renewcommand{\figurename}{Fig.} 
\includegraphics[width=1\textwidth, center]{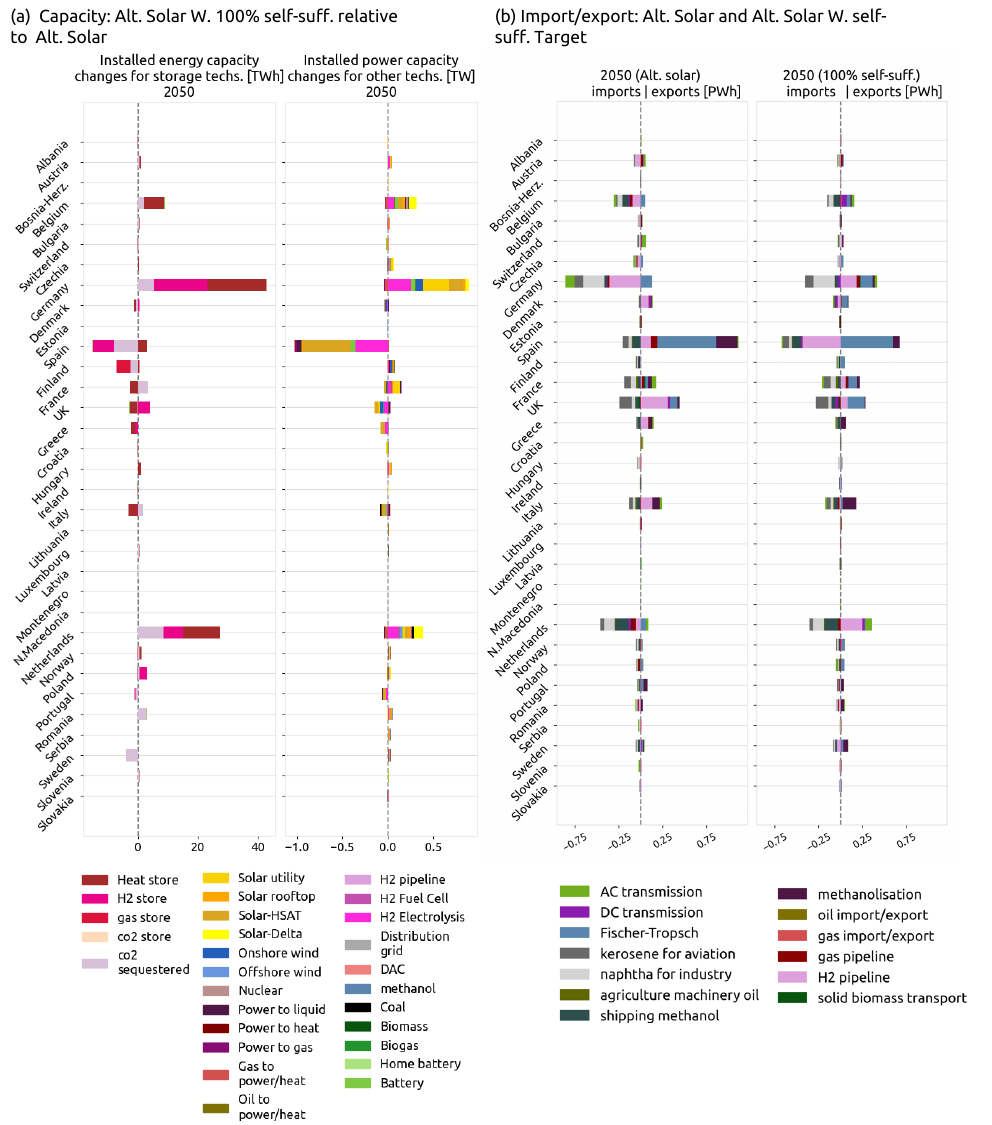}
\caption{Changes in the a) installed capacity of different technologies for the scenario with alternative solar configurations and 100\% self-sufficiency target relative to the scenario with alternative solar configurations in 2050, and b) import and export of different countries for the scenario with alternative solar configurations and the scenario with alternative solar configurations plus 100\% self-sufficiency target in 2050.}
\label{fig:S21}
\end{figure}

\subsection*{S6. Comparison of historical installations for wind and solar in European countries to future needed installations}

\begin{figure}[H]
\renewcommand*{\thefigure}{S\arabic{figure}} \renewcommand{\figurename}{Fig.} 
\includegraphics[width=1\textwidth, center]{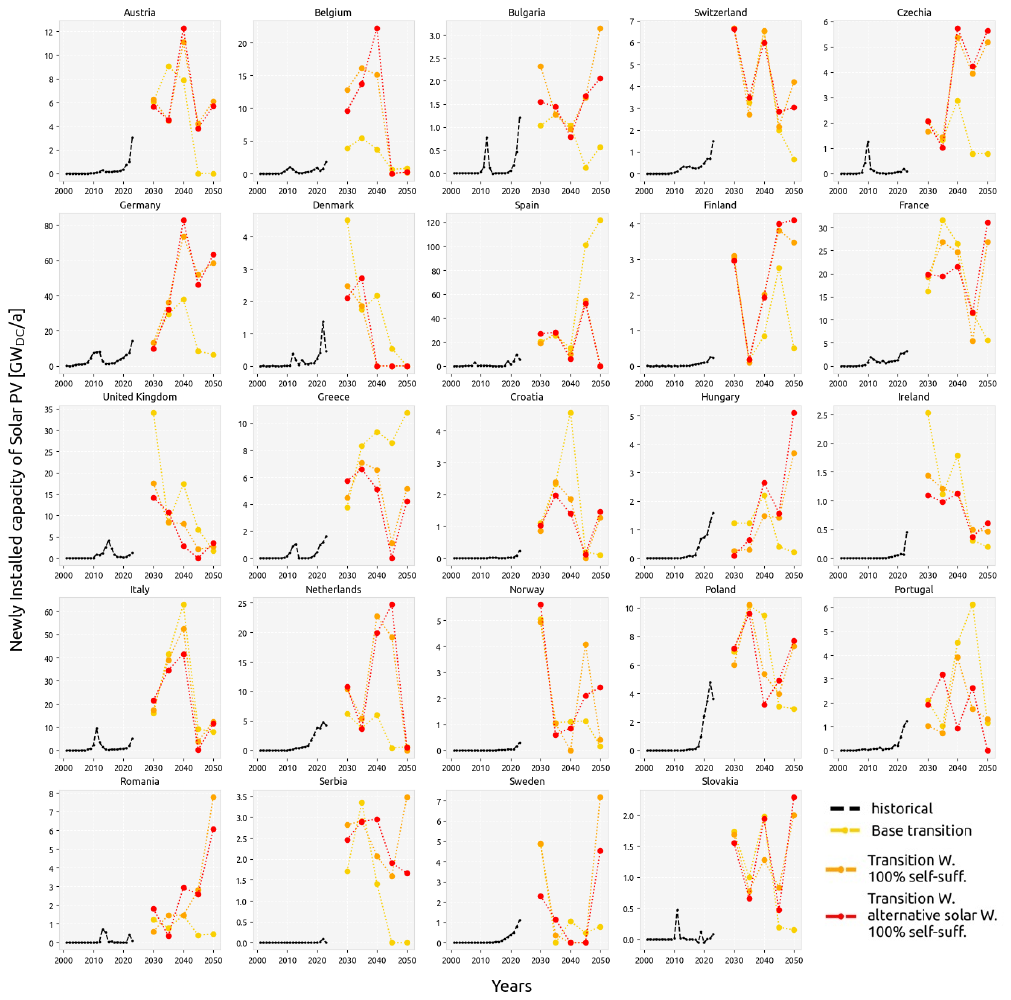}
\caption{Historical (data from IRENA \protect\citeS{IRENA_2022_S}) and future newly installed capacity of solar PV for different European countries modeled under base transition, transition with a 100\% self-sufficiency target, and transition with selected alternative solar configurations under a 100\% self-sufficiency target.}
\label{fig:S23}
\end{figure}

\begin{figure}[H]
\renewcommand*{\thefigure}{S\arabic{figure}} \renewcommand{\figurename}{Fig.} 
\includegraphics[width=1\textwidth, center]{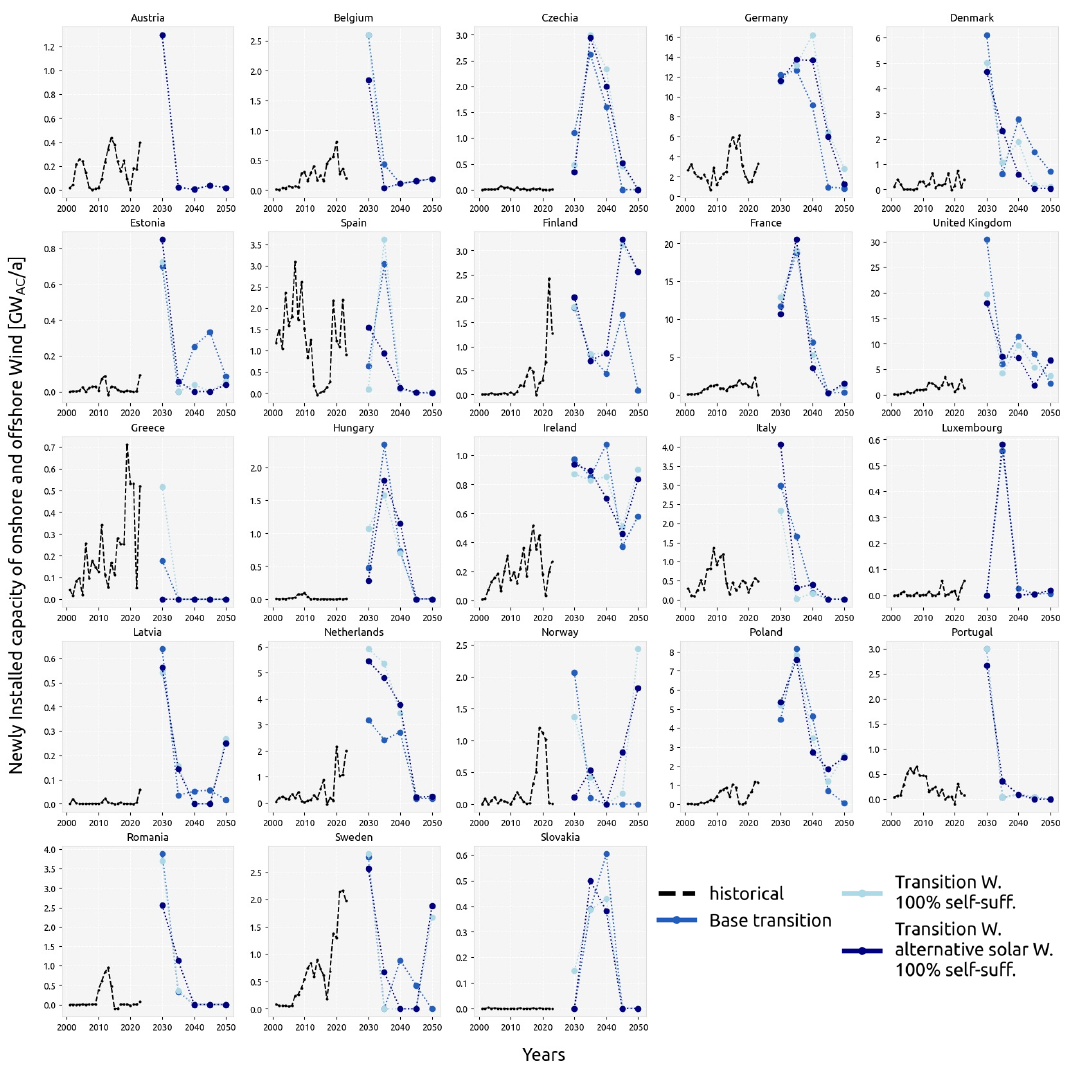}
\caption{Historical (data from IRENA \protect\citeS{IRENA_2022_S}) and future newly installed capacity of onshore and offshore wind for different European countries modeled under base transition, transition with a 100\% self-sufficiency target, and transition with selected alternative solar configurations under a 100\% self-sufficiency target.}
\label{fig:S25}
\end{figure}

\begin{figure}[H]
\renewcommand*{\thefigure}{S\arabic{figure}} \renewcommand{\figurename}{Fig.} 
\includegraphics[width=1\textwidth, center]{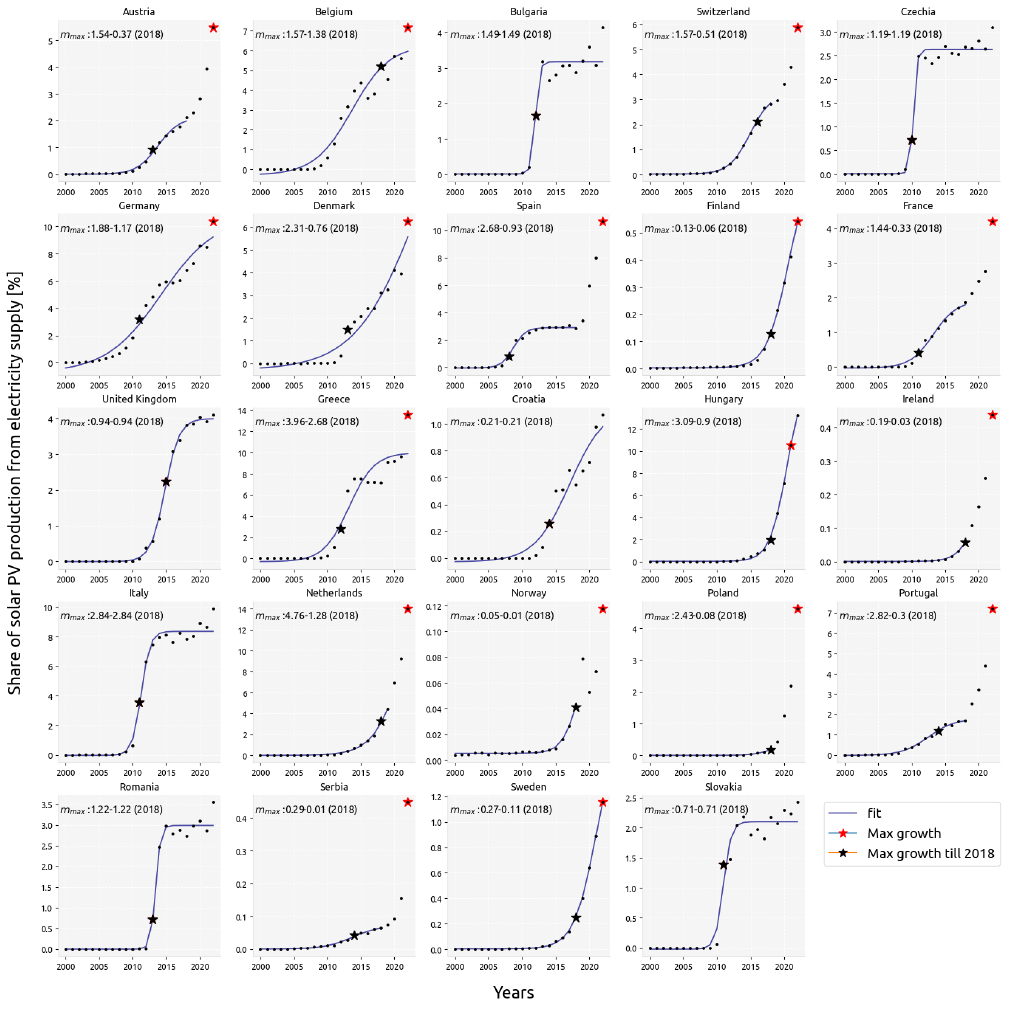}
\caption{Historical (data from IRENA \protect\citeS{IRENA_2022_S}) share of solar PV generation from total electricity generation for different European countries. The blue line shows a fit for the data produced with the logistic function, which has a characteristic S-shaped curve, using the SciPy package (This function is also referred to as the logistic model in the work of Cherp at al. \protect\citeS{cherp2021national_S}). Since many countries already show a second ramp-up period after the tail of the S-curve, a fit cannot be produced for all the data, in which case the fit is produced only for data up to 2018 to match the study by Cherp and co-authors. The black star indicates the maximum growth rate ($m_{max}$) until 2018 and the red star indicates the same for all the data, with both numbers displayed on each figure.  Austria, Belgium, Germany, Spain, and Netherlands are clear examples of countries where the highest growth has taken place very recently (2023), with the value up to five times higher than the maximum growth before 2019. }
\label{fig:S26}
\end{figure}

\begin{figure}[H]
\renewcommand*{\thefigure}{S\arabic{figure}} \renewcommand{\figurename}{Fig.} 
\includegraphics[width=1\textwidth, center]{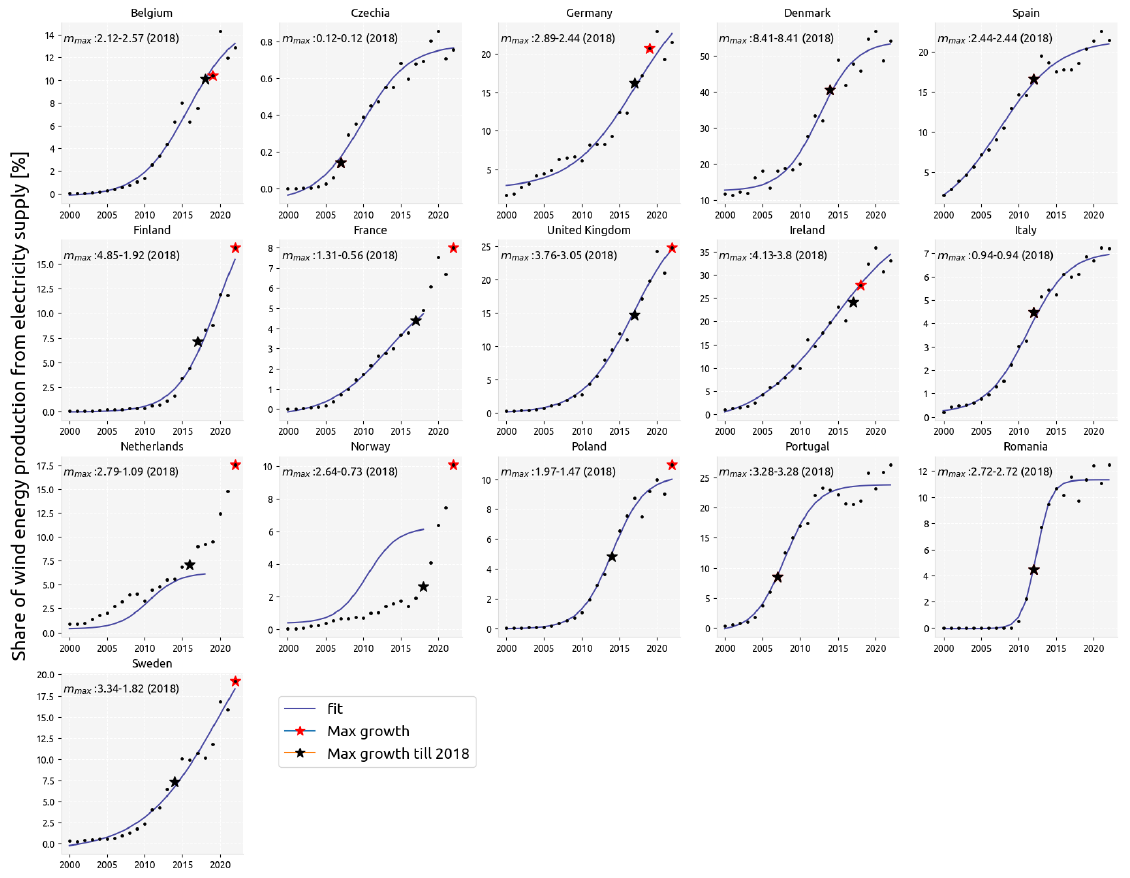}
\caption{Historical (data from IRENA  \protect\citeS{IRENA_2022_S}) share of wind generation (both onshore and offshore) from total electricity generation for different European countries. The blue line shows a fit for the data produced with the logistic function, which has a characteristic S-shaped curve, using the SciPy package. Since some countries already show a second ramp-up period after the tail of the S-curve, a fit cannot be produced for all the data, in which case the fit is produced only for data up to 2018. The black star indicates the maximum growth rate ($m_{max}$) until 2018 and the red star indicates the same for all the data, with both numbers displayed on each figure.  Germany, France, and Netherlands are clear examples of countries where the highest growth has taken place very recently (2023), with the value up to 3.5 times higher than the maximum growth before 2019. }
\label{fig:S27}
\end{figure}

\subsection*{S7. Duration curves for energy sufficiency of select countries during the year}

\begin{figure}[H]
\renewcommand*{\thefigure}{S\arabic{figure}} \renewcommand{\figurename}{Fig.} 
\includegraphics[width=0.9\textwidth,]{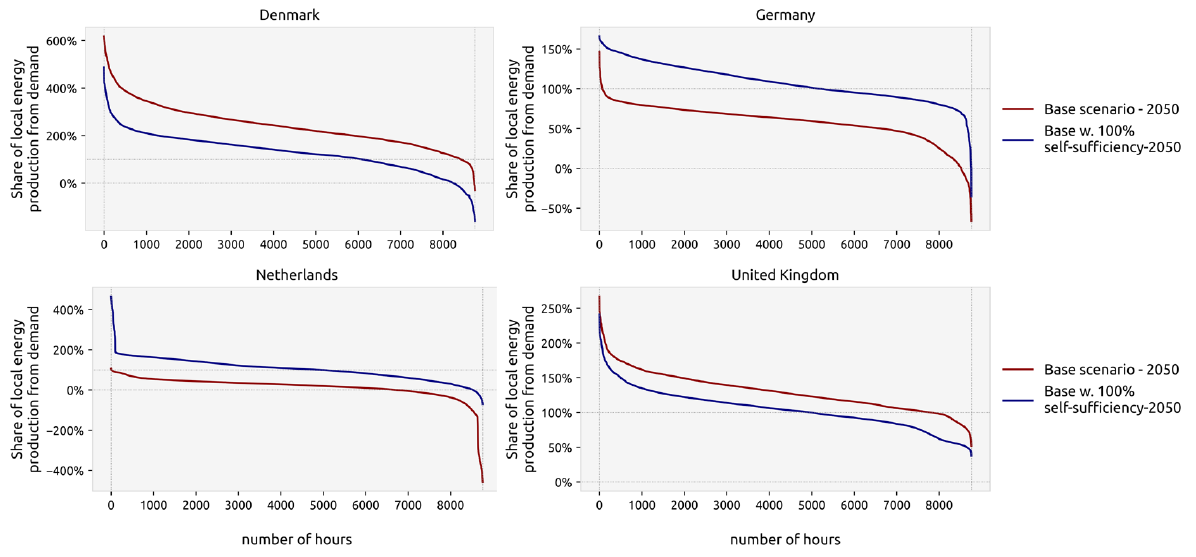}
\caption{Duration curves showing energy self-sufficiency (share of local energy production from demand for all energy carriers) for Germany, Denmark, Netherlands, and the UK for base scenario and base scenario with 100\% self-sufficiency target in 2050. }
\label{fig:S28}
\end{figure}

\subsection*{S8. Energy maps}

\begin{figure}[H]
\renewcommand*{\thefigure}{S\arabic{figure}} \renewcommand{\figurename}{Fig.} 
\includegraphics[width=0.9\textwidth, center]{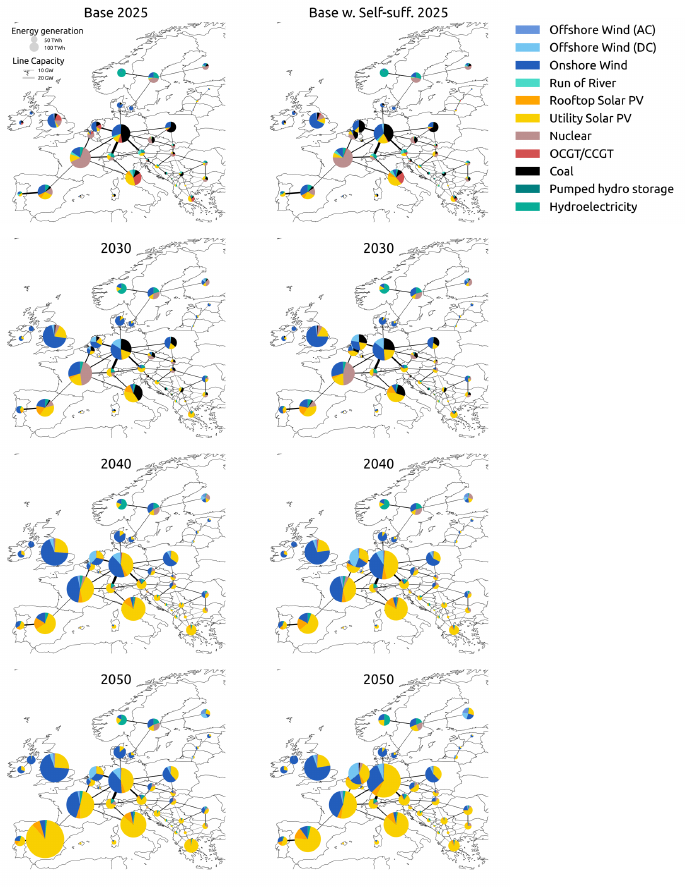}
\caption{Energy generation map for the base scenario (left column) without and (right column) with self-sufficiency targets during the transition showing the share of major technologies in total annual electricity generation. }
\label{fig:S29}
\end{figure}

\begin{figure}[H]
\renewcommand*{\thefigure}{S\arabic{figure}} \renewcommand{\figurename}{Fig.} 
\includegraphics[width=0.9\textwidth, center]{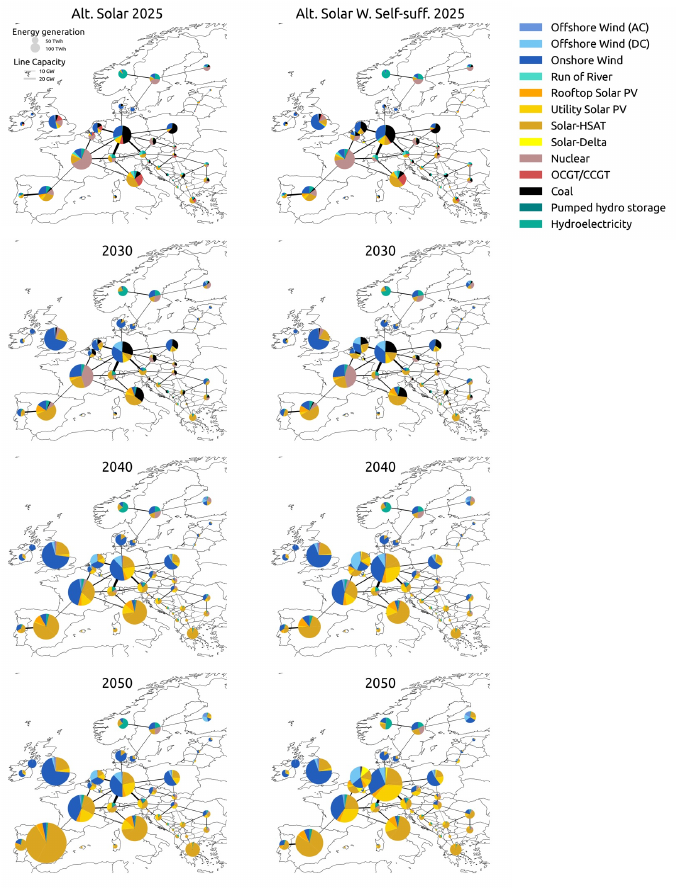}
\caption{Energy generation map for the scenario with alternative solar configurations (left column) without and (right column) with self-sufficiency targets during the transition showing the share of major technologies in total annual electricity generation}
\label{fig:S30}
\end{figure}

\begin{figure}[H]
\renewcommand*{\thefigure}{S\arabic{figure}} \renewcommand{\figurename}{Fig.} 
\includegraphics[width=0.9\textwidth, center]{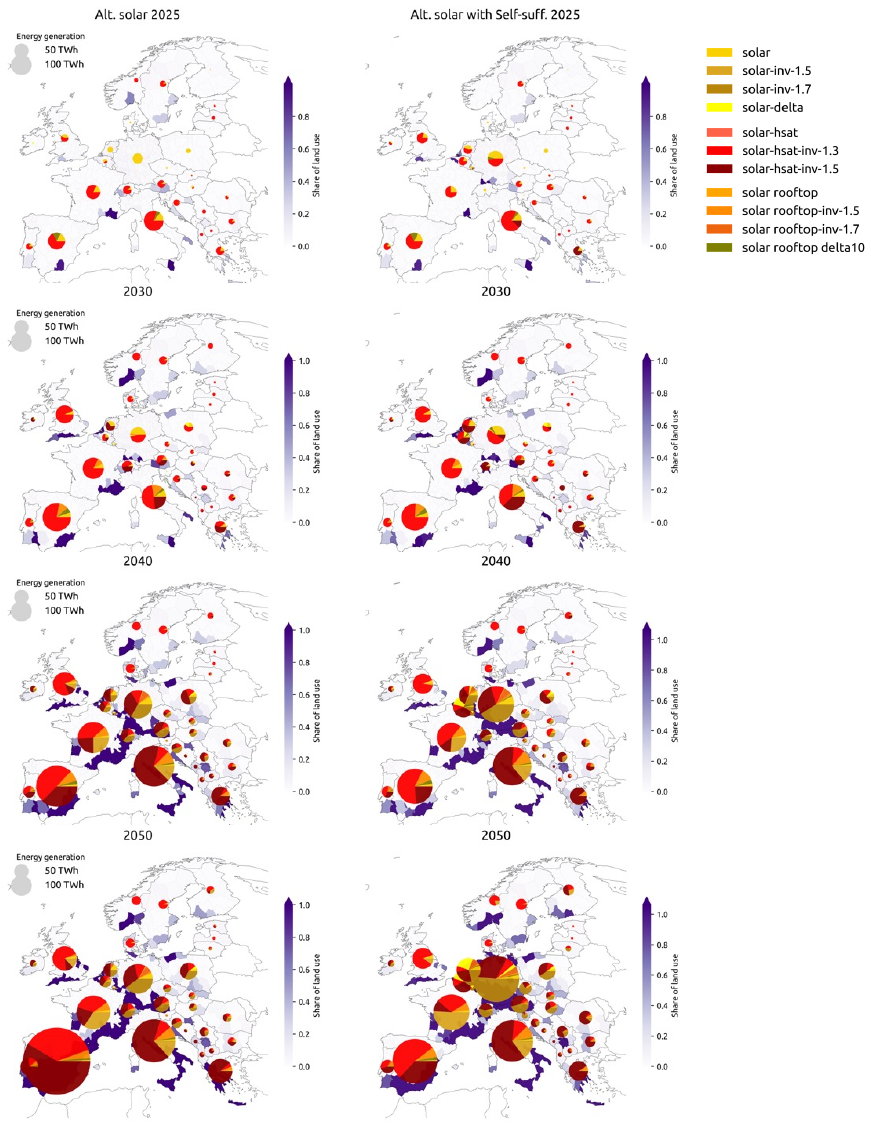}
\caption{Solar generation map by configuration and regional cumulative land-use of solar PV technologies for year years 2025, 2030, and 2040 for (left column) Transition with selected alternative solar configurations, and (right column) Transition with selected alternative solar configurations with self-sufficiency target.}
\label{fig:S31}
\end{figure}

\begin{figure}[H]
\renewcommand*{\thefigure}{S\arabic{figure}} \renewcommand{\figurename}{Fig.} 
\includegraphics[width=0.9\textwidth, center]{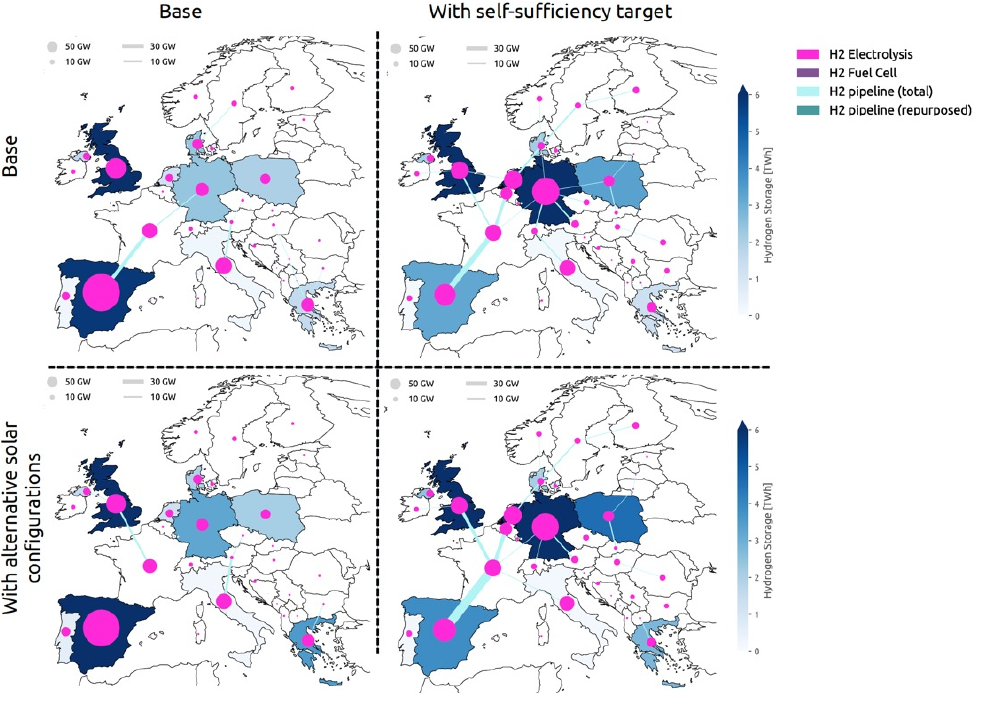}
\caption{Map of energy infrastructure for hydrogen generation and transport for different scenarios in 2050. Addition of alternative solar configurations does not impact the infrastructure layout noticeably. The self-sufficiency requirement causes a large shift of installed capacities from Spain to Germany, Belgium, and Netherlands as they increase their hydrogen production. }
\label{fig:S32}
\end{figure}

\begin{figure}[H]
\renewcommand*{\thefigure}{S\arabic{figure}} \renewcommand{\figurename}{Fig.} 
\includegraphics[width=0.9\textwidth, center]{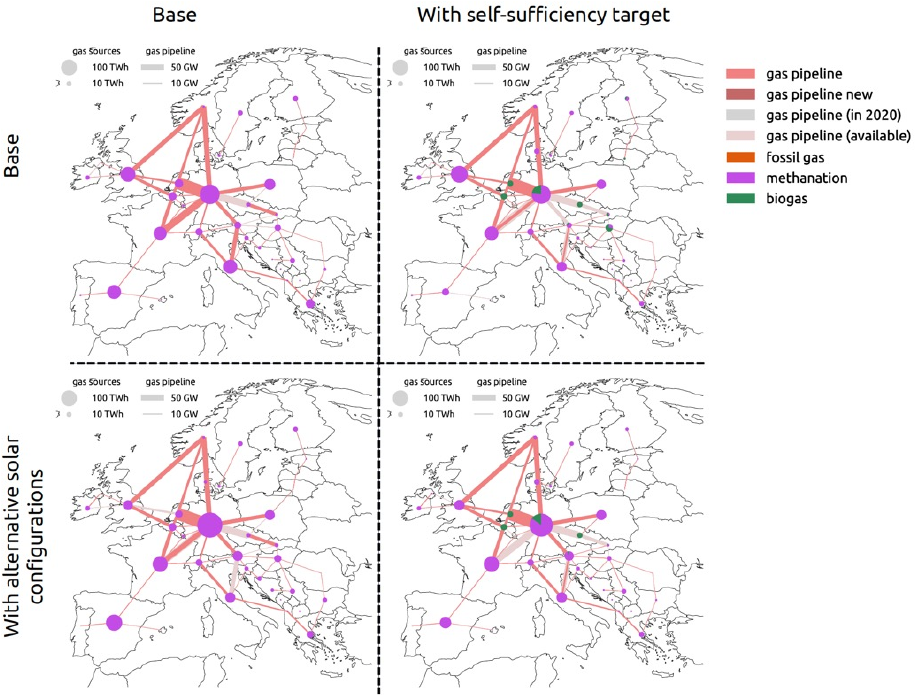}
\caption{Map of energy infrastructure for methane generation and transport for different scenarios in 2050. Addition of alternative solar configurations does not impact the infrastructure layout noticeably. The self-sufficiency requirement triggers the installation of biomass plants in Germany, Netherlands, and Belgium, as they reduce their import of hydrogen that is required for methanation (see Fig. \ref{fig:S17}).}
\label{fig:S33}
\end{figure}

\begin{figure}[H]
\renewcommand*{\thefigure}{S\arabic{figure}} \renewcommand{\figurename}{Fig.} 
\includegraphics[width=0.9\textwidth, center]{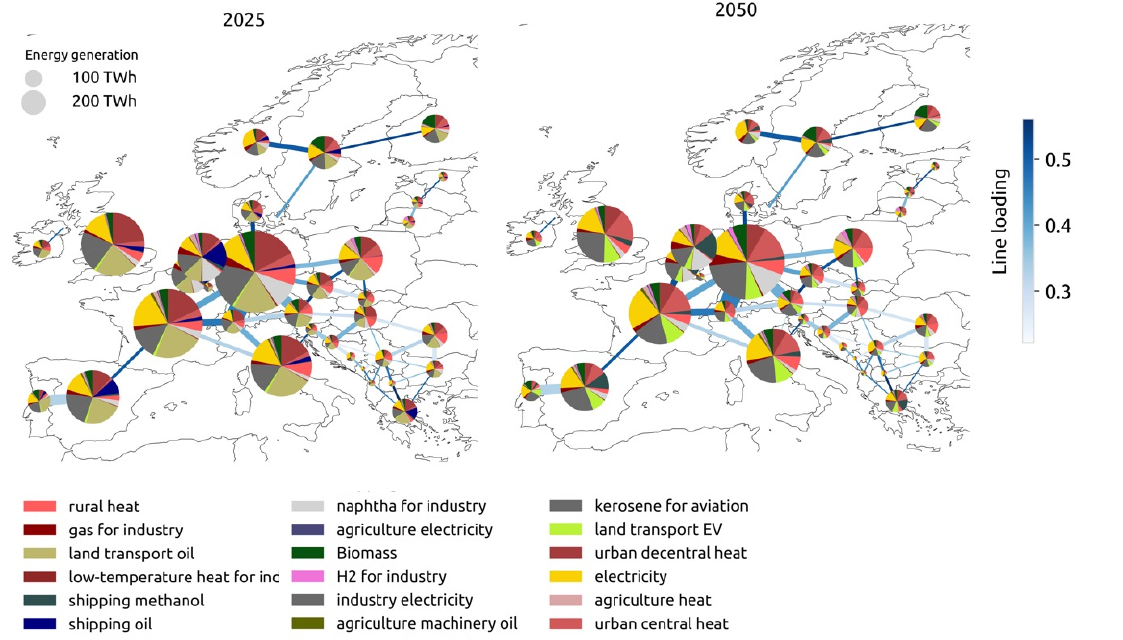}
\caption{Total energy demand by sector for years 2025 and 2050. Notice the transition of shipping fuel from oil to methanol and land transport from oil to electricity.}
\label{fig:S34}
\end{figure}

\clearpage
\newpage
\fontsize{10}{12}\selectfont
 \bibliographystyleS{naturemag}
\bibliographyS{Bibliography.bib}

\clearpage
\tableofcontents

\end{document}